\documentclass[iop,apj,tighten]{emulateapj}
\usepackage{apjfonts} 
\usepackage{bm}
\usepackage{amstext,mathtools}
\usepackage[breaklinks,colorlinks,citecolor=blue,linkcolor=magenta]{hyperref} 
\usepackage[all]{hypcap} 
\usepackage[usenames,dvipsnames]{color}

\def\be{\begin{equation}}
\def\ee{\end{equation}}
\def\ben{\begin{eqnarray}}
\def\een{\end{eqnarray}}

\DeclareMathAccent{\dot}    {\mathalpha}{operators}{'137} 
\DeclareMathAccent{\ddot}    {\mathalpha}{operators}{'177} 

\newcommand{\Msun}{\mathrm{M}_\odot}

\shorttitle{Universality in the PBH merger distribution}
\shortauthors{Kocsis, Suyama, Tanaka, Yokoyama}

\begin{document}


\title{
Hidden universality in the merger rate distribution in the primordial black hole scenario
}

\author{
Bence Kocsis\altaffilmark{1},
Teruaki Suyama\altaffilmark{2},
Takahiro Tanaka\altaffilmark{3,4},
and Shuichiro Yokoyama\altaffilmark{5,6}
}

 \affil{$^1$Institute of Physics, E\"otv\"os University, P\'azm\'any P. s. 1/A, Budapest, 1117, Hungary;}

 \affil{$^2$   Research Center for the Early Universe (RESCEU), Graduate School
   of Science,\\ The University of Tokyo, Tokyo 113-0033, Japan
 }

 \affil{$^3$   Department of Physics, Kyoto University, Kyoto 606-8502, Japan
 }
 
  \affil{$^4$   
   Center for Gravitational Physics, Yukawa Institute for Theoretical Physics,
Kyoto University, Kyoto 606-8502, Japan
 }

 \affil{$^5$   
   Department of Physics, Rikkyo University, Tokyo 171-8501, Japan
 }

 \affil{$^6$   
   Kavli IPMU (WPI), UTIAS, The University of Tokyo,Kashiwa, Chiba 277-8583, Japan
 }

\begin{abstract}
It has been proposed that primordial black holes (PBHs) form binaries in the radiation dominated era.
Once formed, some fraction of them may merge within the age of the Universe by gravitational radiation reaction.
We investigate the merger rate of the PBH binaries
when the PBHs have a distribution of masses around ${\cal O}(10) \Msun$,
which is a generalization of the previous studies where the PBHs are assumed to have the same mass.
After deriving a formula for the merger time probability distribution in the PBH mass plane,
we evaluate it under two different approximations.
We identify a quantity constructed from the mass-distribution of the merger rate density per unit cosmic time and 
comoving volume 
$\mathcal{R}(m_1,m_2)$, $\alpha = -{(m_1+m_2)}^2\partial^2 \ln\mathcal{R}/\partial m_1\partial m_2 $, 
which universally satisfies $0.97 \lesssim \alpha \lesssim 1.05$ for all binary masses independently of the PBH mass function.
This result suggests that the measurement of this quantity is useful for testing the PBH scenario.
\end{abstract}


 \maketitle

\section{Introduction}
Recent detections of gravitational wave events (GW150914, LVT151012, GW151226, GW170104, GW170608,
and GW170814)
by the LIGO-Virgo collaboration \citep{Abbott:2016blz, TheLIGOScientific:2016pea, Abbott:2017vtc, Abbott:2017oio, Abbott:2017gyy}
revealed the existence of binary black holes (BHs) in the mass range 
$8$--$35~\Msun$. 
These observations clearly demonstrate that there are numerous BH-BH binaries in the Universe
that have previously eluded the scrutiny by astronomers.
The origin of such heavy BHs and the formation of close binary BHs which merge within the age of the Universe are widely debated.
Various astrophysical scenarios for the explanations of the gravitational wave events
are summarized, for instance, in \citet{TheLIGOScientific:2016htt} and \citet{Miller:2016krr}.

Although only five robustly identified BH-BH binary mergers with GW detections 
have been reported so far, merger rates are constrained to within $12$--$240\,{\rm Gpc}^{-3}{\rm yr}^{-1}$ \citep{Abbott:2017vtc}. 
With the further improvement of GW detectors, we will soon enter the era 
of {\it black hole rush} where a large number of BH-BH binaries
are detected with their masses, spins, and locations determined.
Those data will serve us important clues to clarify the origin of binary BHs as well 
as the formation mechanism of the binaries.
Clearly, investigations of how various astrophysical scenario producing merging BH binaries can be distinguished by 
observations will become a fundamentally important topic.

Recently, a collaboration including three of the authors, \citet{Sasaki:2016jop} pointed out that the GW event GW150914
could be merger events of two primordial black holes (PBHs) 
based on earlier studies \citep{Nakamura:1997sm,Ioka:1998nz}. 
In \citet{Nakamura:1997sm} and \citet{Ioka:1998nz}, the formation
mechanism of the PBH binaries was proposed and a connection between the PBH binaries and 
the gravitational wave events from the merger of binary PBHs was given\footnote{
There are other papers in which potential detection of PBHs by LIGO was claimed
\citep{Bird:2016dcv, Clesse:2016vqa, Kashlinsky:2016sdv}. 
The binary formation path is different from that in \citet{Sasaki:2016jop}.}.
PBHs stand for BHs that formed in the very early Universe much before the epoch of the
matter radiation equality \citep{Carr:1974nx}.
For instance, in the well-studied scenario, PBHs form from rare high peaks 
of the primordial density inhomogeneities whose amplitudes are much
larger than the standard deviation.
In this case, the PBH mass is given by the total energy contained in the
Hubble horizon at the formation time,
\be
m_{\rm BH} =\gamma \frac{4\pi}{3}\rho H^{-3}\approx 30~\Msun ~\left( \frac{\gamma}{0.2} \right)
{\left( \frac{T}{30~{\rm MeV}} \right)}^{-2},
\ee
where $T$ is the temperature of radiation and $\gamma={\cal O}(1)$ depends on the details
of the BH formation.
Analytic estimates give $\gamma=3^{-3/2} \approx 0.2$ \citep{Carr:1975qj}.
Other mechanisms of the PBH production are summarized by \citet{Carr:2005zd}. 
After having formed in the very early Universe, PBHs stay on the expansion flow of the Universe.
Even when PBHs are randomly distributed in space without being clustered,
there is a small but non-vanishing probability that two neighboring
PBHs happen to be much closer than the mean distance.
Such PBHs, being initially on the cosmic expansion flow,
eventually start to come closer influenced by their mutual gravity when the cosmic expansion rate becomes too low to separate them apart.
As was shown by \citet{Nakamura:1997sm}, a direct collision is avoided by the tidal effect of other PBHs in their vicinity, which leads to the formation of 
a PBH binary with a large eccentricity. 
Further \citet{Ali-Haimoud:2017rtz} have recently shown that the tidal field of halos and interactions with other PBHs, 
as well as dynamical friction by unbound dark matter particles, do not affect PBH binaries significantly.
Highly eccentric PBH binaries radiate GWs efficiently and 
a fraction of them can merge within 14 billion years.

In \citet{Sasaki:2016jop}, under the approximation that all PBHs have the same mass of $30~\Msun$,
it was shown that the expected event rate of the PBH binary mergers is consistent with
the one determined by the LIGO-Virgo collaboration after the announcement of GW150914 \citep{Abbott:2016nhf},
if the fraction of cold dark matter in PBHs is about $10^{-3}$.
This fraction is consistent with existing observational upper limits 
\citep{Gaggero:2016dpq,
Horowitz:2016lib, Brandt:2016,Koushiappas:2017chw,Inoue:2017csr, Green:2017qoa,Matsumoto:2017adh,Carr:2017jsz,Poulin:2017bwe}.
So far, the PBH scenario proposed by \citet{Sasaki:2016jop} is successful in explaining
the LIGO event GW150914.

In the next decades, many more BH binaries will be detected, which will deliver 
fruitful statistical information on the merger rates in the two-dimensional BH mass plane $(m_1,m_2)$
(see \citealt{TheLIGOScientific:2016pea,OLeary:2016ayz,Mandel:2016prl,Zevin:2017evb,Kovetz:2017,Fishbach:2017zga,Gondanetal2017}).
Purpose of the present paper is to examine if the mass distribution can be used observationally to test the PBH scenario. 
The currently announced five robust merger events show some scatter in
the BH mass as $(m_1,m_2)=(36^{+5}_{-4},29^{+4}_{-4})$ for GW150914, 
$(14.2^{+8.3}_{-3.7},7.5^{+2.3}_{-2.3})$ for GW151226, $(31.2^{+8.4}_{-6.0},19.4^{+5.3}_{-5.9})$ for GW170104,
$(12^{+7}_{-2},7^{+2}_{-2})$ for GW170608,
and $(30.5^{+5.7}_{-3.0},25.3^{+2.8}_{-4.2})$ for GW170814 in units of solar mass ($90\%$ credible intervals)
\citep{Abbott:2016blz, TheLIGOScientific:2016pea, Abbott:2017vtc, Abbott:2017oio, Abbott:2017gyy}.
In this paper, we estimate the merger rate density in the $m_1-m_2$ plane 
predicted by the PBH scenario.
We extend the formalism 
of previous studies \citep{Nakamura:1997sm, Ioka:1998nz, Sasaki:2016jop} to compute the merger event rate 
to the case in which the PBH mass function is not restricted to a single-mass but it 
extends over a mass range between 
$m_{\min}$ and $m_{\max}$ with $m_{\max}/m_{\min}\lesssim 10$ \footnote{
Recently, such an extension has also been done in \citep{Raidal:2017mfl}.
Our study differs from \citep{Raidal:2017mfl} in that our primary purpose is to investigate 
the universal feature of the merger-rate distribution
that is insensitive to the PBH mass function.
}.
We assume that the PBH mass function does not extend over many orders of magnitude 
since in that case the dynamics may not be accurately captured by 
the simple physical processes adopted by \citet{Nakamura:1997sm, Ioka:1998nz, Sasaki:2016jop}. 
Quite interestingly, we find that the merger rate distribution in this case depends on the mass of the BH binary
in a specific way and a quantity constructed from the mass-distribution of the merger rate density per unit time and volume $\mathcal{R}(m_1,m_2)$,
\be\label{alphadef}
\alpha = -{(m_1+m_2)}^2 \partial^2 \ln\mathcal{R}/\partial m_1\partial m_2,
\ee
is insensitive to the PBH mass function.
This distinct feature is advantageous since there is no theoretically 
tight constraint on the shape of the PBH mass function. 
Identifying the information in the merger rate density which is insensitive to the BH mass function may be used to discriminate different formation channels \citep{OLeary:2016ayz,Kovetz:2017, Zevin:2017evb, Gondanetal2017}. 
This information may be used to obtain the probability of mergers for given BH masses, 
$P_{\rm intr}(m_1,m_2)$ (defined by Eq.~(\ref{obs-P}) below) which is essential in measuring 
the underlying BH mass function $f(m)$ itself.

Before closing this section, in Table~\ref{def-table} we list definitions of important symbols that are used in this paper.
\begin{table}[t]
	\begin{tabular}{|l|c|} \hline
    Symbols & Meaning \\ \hline
	$m_1, ~m_2$ & Mass of the individual PBHs in binary \\
    $m_{\rm t}$ & Total mass $m_1+m_2$ \\
    $n_{\rm BH}$ & Comoving PBH number density \\
    $f_{\rm PBH}$ & Fraction of PBHs in dark matter \\
    $f (m)$ & PBH mass function with normalization condition (\ref{pbh-mf}) \\
    ${\cal R} (m_1,m_2,t)$ & Merger rate density per unit cosmic time $t$ and comoving volume \\
    $P_{\rm intr}(m_1,m_2,t)$ & Intrinsic merger rate density defined by Eq.~(\ref{obs-P}) \\
    $\alpha$ & Universal rate exponent defined by Eq.~(\ref{alphadef}) \\
    $D$ & Physical distance between PBHs that form a binary \\
    $M_i$ & Mass of $i$-th outer PBH \\
    $D_i$ & Physical distance to $i$-th outer PBH (see Fig.~\ref{fig-tidal}) \\
    $y_i$ & Comoving distance to $i$-th outer PBH \\
    $\theta_i$ & Angle (see Fig.~\ref{fig-tidal}) \\
    ${\vec e}_i$ & Vector (see Fig.~\ref{fig-tidal} and Eq.~(\ref{unit-vector-ei})) \\
    $x$ & Comoving distance between PBHs that form a binary \\
    $x_{\rm max}$ & Maximum value of $x$ to form binary (see Eq.~(\ref{eq-xmax})) \\
    $z_{\rm dec}$ & Redshift when PBHs form a binary \\
    $t_{\rm dec}$ & Cosmic time corresponding to $z_{\rm dec}$ \\
    $A$ & Defined by Eq.~(\ref{major-axis}) \\
    $a$ & Initial major-axis of PBH binary \\
    $a_{\rm max}$ & $a_{\rm max}=x_{\rm max}/(1+z_{\rm eq})$ \\
    $e$ & Initial eccentricity of PBH binary \\
    $e_{\rm m}$ & Maximum eccentricity given by Eq.~(\ref{def-em}) \\
    $\zeta$ & Length of ${\vec \zeta}$ defined by Eq.~(\ref{eccen-e}) \\
    $F(x,\zeta)$ & Probability density of $(x,\zeta)$ (see Eq.~(\ref{F-x-zeta})) \\
    $t$ & Cosmic time when PBH binary merges \\
    $\tau$ & Time delay between binary formation and merger: $\tau = t-t_{\rm dec}$ \\
    $\beta$ & $\sin (2\theta_1)$ (see Eq.~(\ref{only-nearest-g}) and below it) \\
    $K$ & Dimensionless quantity defined by Eq.~(\ref{definition-of-K}) \\
    $m_c$ & Defined by Eq.~(\ref{definition-of-mc}) \\
    $G(x)$ & Defined by Eq.~(\ref{definition-of-G(x)}) \\
    $m_{\rm min}$ & Minimum PBH mass of flat mass function (\ref{flatmf}) \\
    $m_{\rm max}$ & Maximum PBH mass of flat mass function (\ref{flatmf}) \\
    ${\tilde \zeta}$ & Defined by ${\tilde \zeta}=(m_{\rm t}/m_{\rm max}) \zeta$ (See \ref{case2}) \\
    $\sigma, {\tilde \sigma}$ & $\sigma$ defined by Eq.~(\ref{flat-variance}) and ${\tilde \sigma}=(m_{\rm t}/m_{\rm max}) \sigma$ \\
    $\xi$ & Fitting parameter appearing in Eq.~(\ref{P-ansatz}) \\
    $\nu$ & Dimensionless quantity defined by Eq.~(\ref{definition-nu}) \\
    $w_{\rm m}$ & Defined by Eq.~(\ref{definition-zm}) \\
    \hline 
    \end{tabular}
	\caption{Definitions of important symbols that are used in this paper.}
	\label{def-table}
\end{table}

The paper is organized as follows.
We first develop a formalism to compute the event rate in the PBH scenario
which can be applied to the case of a non-monochromatic\footnote{By ``monochromatic mass function'' we refer to a population in which all PBHs have the same mass.} mass function.
Then, we apply the derived formula to evaluate the mass-dependence of the merger rate 
in the $(m_1,m_2)$ BH mass plane and show that the special quantity constructed out of the 
event rate density becomes almost independent of the PBH mass function. 

\section{Formation of binary PBHs}
In this section, we derive a formula of the merger rate density as a function of the 
masses of two BHs comprising the binary.

\subsection{Formation and mass function of PBHs}
There are several mechanisms to form PBHs \citep{Carr:2005zd}.
Among them, the most natural and widely investigated mechanism is the direct gravitational
collapse of the primordial density perturbation in the radiation dominated Universe. 
In this scenario, when an overdense region containing an extremely high density peak in which the perturbation amplitude 
is greater than $\delta_{\rm th} ={\cal O}(1)$ reenters the Hubble horizon,
that region directly collapses to a BH (for the estimation of $\delta_{\rm th}$, 
see \citep{Carr:1975qj, Harada:2013epa}).
Crudely speaking, all the energy inside the Hubble horizon at the time of BH formation turns into the BH.
This picture enables to relate the BH mass to the comoving wavenumber $k$ of the primordial density perturbation as 
\be
m_{\rm PBH}\sim 20~\Msun~{\left( \frac{k}{1~{\rm pc}^{-1}} \right)}^{-2}. \label{mass-k}
\ee
There are no direct observational constraints on the probability distribution
of density perturbations on such small scales.

Although Eq.~(\ref{mass-k}) gives us a simple and approximate estimate of the PBH mass in terms of $k$,
the relation (\ref{mass-k}) is not precisely correct since the PBH mass also depends
on the amplitude of the density perturbation.
Deviation of the actual PBH mass from the horizon mass becomes significant as the amplitude
of the density perturbation approaches $\delta_{\rm th}$ \citep{Choptuik:1992jv,Niemeyer:1997mt}.
Thus, even if the spectrum of the primordial density perturbation is monochromatic,
the resulting PBH mass function is not monochromatic \citep{Yokoyama:1998xd}.
Furthermore, the power spectrum of the primordial density perturbations needs not be monochromatic.
In the paradigm of the standard inflationary cosmology, the primordial density perturbations
are produced in the inflationary era preceding the radiation dominated era.
Several inflationary models have been proposed to date with different predictions for the power spectral shape of the primordial density perturbation 
which lead to different PBH numbers and mass functions (see \citep{Carr:2016drx} and references therein).
To a varying degree, these models predict a non-monochromatic power spectrum.
Thus, the PBH mass function is generally not concentrated on a single mass.

The PBH mass function is determined once the inflation model is fixed and the power
spectrum of the primordial density perturbation is computed\footnote{In addition, 
non-Gaussianity of the primordial density perturbation also affects the PBH mass function \citep{Byrnes:2012yx,Young:2013oia}.}.
Since there is no fiducial inflation model producing PBHs and different models predict different
PBH mass functions, we do not restrict our analysis to any particular PBH mass function.
As mentioned earlier, our only requirement is that it is confined to the mass range 
$m_{\max}/m_{\min}\lesssim 10$.
The case where the PBH mass function is extended over many orders of magnitude
requires a separate analysis, which is beyond the scope of this paper.

In addition to the mass function, the spatial distribution of PBHs also affects the probability of binary formation.
In this study, for simplicity we assume that the distribution of PBHs at their birth is 
statistically uniform and random in space.
However, we also have to keep in mind that primordial 
clustering of PBHs is also possible and could be 
an important factor to enhance the merger event rate for a fixed mass fraction of PBHs. 
We define the PBH mass function $f(m)$ such that $f(m) dm$ is the 
probability that a randomly chosen PBH has mass in $(m,m+dm)$.
Thus, $f(m)$ is normalized as
\be
\int_{m_{\rm min}}^{m_{\rm max}} f(m)dm=1. \label{pbh-mf}
\ee
We denote the comoving PBH number density as $n_{\rm BH}$.
The mean comoving separation between two neighboring BHs is thus given by $n_{\rm BH}^{-1/3}$.

Before closing this subsection, it is important to mention that
we do not consider the mass growth of the PBHs following their initial formation.
The mass change due to accretion is negligible when PBH is in environments similar to
the cosmic average density \citep{Carr:1974nx, Custodio:1998pv, Ali-Haimoud:2016mbv}.
This may not be true for PBHs residing in high density regions of galaxies such as molecular clouds, accretion disks, or stellar interiors.
However, since the majority of PBHs are expected to remain mostly in low density regions 
such as dark matter halos, we ignore the mass growth of PBHs.

\subsection{Major axis and eccentricity of a binary}
\label{subsec:Mm}
\begin{figure}
\centering
\includegraphics[width=\columnwidth]{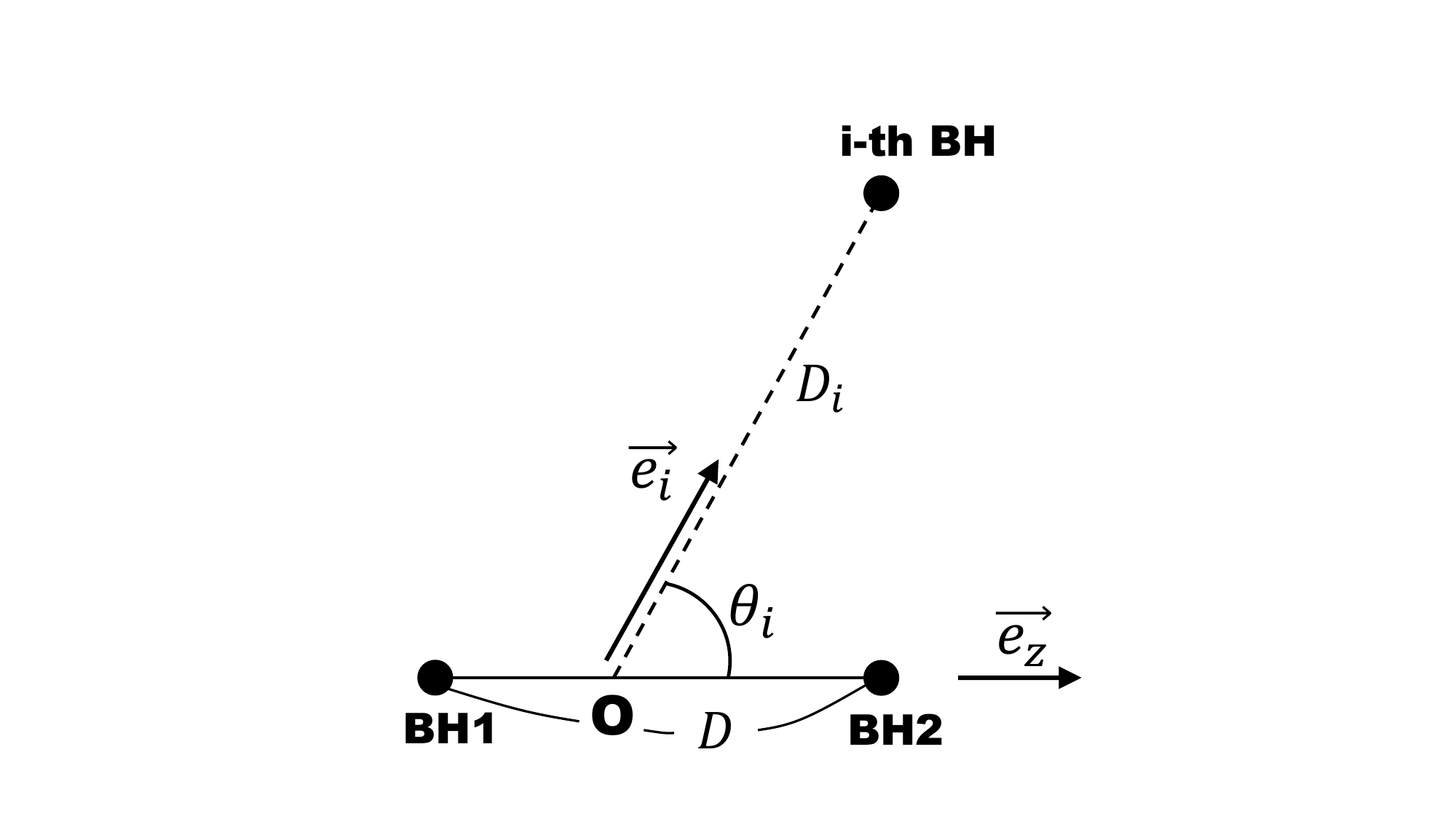}
\caption{ \label{fig-tidal}Schematic picture showing the spatial configuration of BHs.}
\end{figure}

Just after PBHs are formed in the early Universe, 
they are typically separated by super-Hubble distances.
Apart from a possible peculiar velocity, 
each PBH is attached to the flow of the cosmic expansion.
Let us denote the mass of a randomly selected PBH by $m_1$, and the mass of and the comoving distance to the closest PBH by $m_2$ and $x$, respectively.
Denoting the physical distance between the two BHs by $D$ (see Fig.~\ref{fig-tidal}), the gravitational force is given by
$Gm_1 m_2/D^2$.
Ignoring for the moment the subdominant effects of the other remote BHs and the initial peculiar velocity and
assuming that the above gravitational force is the only dynamical effect acting on each 
BH\footnote{In particular, we neglect the gravitational pull of the background density inhomogeneities and the forces 
that arise due to anisotropic accretion from the background density. We will discuss these assumptions below.}, 
the BHs attract each other and
collide within the free-fall time given by
\be
t_{\rm ff} = D^{3/2}/\sqrt{Gm_{\rm t}},~~~~~m_{\rm t} \equiv m_1+m_2.
\ee
In reality, the space is expanding, and the BHs will be distanced if the space expands 
by ${\cal O}(1)$ or more within the free-fall time.
Conversely, if the free-fall time is shorter than the Hubble time $1/H$,
then the two BHs become gravitationally bound and eventually collide.
Since the free-fall time and the Hubble time respectively scale as ${({\rm scale ~factor})}^{3/2}$
and (scale factor)$^{2}$ during the radiation dominated era, the Hubble time
may eventually exceed the free-fall time in the radiation dominated era even if
the BHs are initially on the cosmic expansion flow \citep{Nakamura:1997sm}.
The condition for forming the bound system can be written as
\be
\frac{1}{\sqrt{Gm_{\rm t}}} {\left( \frac{x}{1+z} \right)}^{3/2} < \frac{1}{H(z)},
\ee
where $z$ is the cosmological redshift.
Using the Friedmann equation for a flat cosmology and neglecting factors of order unity, this condition can be rewritten as
\be
m_{\rm t} > \rho (z) \frac{x^3}{{(1+z)}^3}, \label{bound-condition}
\ee
where $\rho (z)$ is the background energy density.
From this expression, we can give another but equivalent physical interpretation 
to the criterion for forming the gravitationally bound state.
The left hand side is the total mass of the two BHs, and the right hand side is the
total mass of whatever matter component that dominates the background Universe.
Thus, the condition for two BHs to become gravitationally bound 
is equivalent to the condition for the total energy $m_{\rm t}$ to exceed the background energy contained in the comoving volume to the nearest PBH $x^3$.

In the radiation dominated era, the energy density of radiation can be written as
\be
\rho (z) \approx \rho_{c,0} \frac{{(1+z)}^4}{1+z_{\rm eq}} \Omega_m,
\ee
where $z_{\rm eq}$ is the redshift at the time of matter-radiation equality,
$\rho_{c,0}$ and $\Omega_m$ respectively represent a critical density and a density parameter of the non-relativistic matter at the present,
and the right-hand side in Eq.~(\ref{bound-condition}) decreases in time.
Then, if $x$ is smaller than $x_{\rm max}$ given by
\be
x_{\rm max}={\left( \frac{m_{\rm t}}{\rho_{c,0} \Omega_m} \right)}^{1/3}, \label{eq-xmax}
\ee
Eq.~(\ref{bound-condition}) becomes satisfied at $z=z_{\rm dec} > z_{\rm eq}$,
where $z_{\rm dec}$ is given by
\be
1+z_{\rm dec}= (1+z_{\rm eq}) {\left( \frac{x_{\rm max}}{x} \right)}^3. \label{decoupling-redshift}
\ee
The physical distance of the BH pair at the time of decoupling time,
which becomes the semimajor axis of the resultant binary,
is given by
\be
a=\frac{1}{1+z_{\rm dec}}x = Ax^4,~~~~~A \equiv \frac{1}{1+z_{\rm eq}} \frac{1}{x_{\rm max}^3} =
\frac{1}{1+z_{\rm eq}} \frac{\rho_{c,0} \Omega_m}{m_{\rm t}}. \label{major-axis}
\ee
Since the BH pair forms only for $x < x_{\rm max}$, there is an upper bound
on $a$ as $a < a_{\rm max} =x_{\rm max}/(1+z_{\rm eq})$.

If there is no force other than the gravitational force from the neighboring BHs, 
and the initial peculiar velocities vanish,
such two BHs come closer by moving on the same straight line and end up with a head-on collision.
However, in reality, there are other remote BHs surrounding the BHs in pair, and they 
exert a torque during the infall motion of the BHs in pair.
As a result, the BH pair acquires an angular momentum, and the head-on collision is circumvented.
The torque exerted by the $i$-th distant BH to the lowest order in the distance $D_i$ to the $i$-th BH 
is given by
\be
N_i =\frac{3GM_i}{2D_i^3} \sin (2\theta_i) \frac{m_1 m_2}{m_{\rm t}} D^2,
\ee
where $D$ is the physical distance between BH1 and BH2 (see Fig.\ref{fig-tidal}), 
$M_i$ is the mass of the $i$-th perturber BH, and $\theta_i$ is the angle between a line connecting two BHs in pair 
and a line connecting $i$-th BH and a center of mass of the BH pair (see Fig.~\ref{fig-tidal}).
Thus, the angular momentum generated by this torque throughout the free fall becomes
\be
J_i \simeq N_i t_{\rm ff}.
\ee
Taking the direction of the torque exerted by each BH into account, 
the total angular momentum that the BH pair acquires is given by 
\be
{\vec J}
=\frac{3}{2} t_{\rm ff} \frac{Gm_1 m_2}{m_{\rm t}} D^2 \sum_{i=1}^N \frac{M_i}{D_i^3} 
\sin (2\theta_i) \frac{({\vec e_z} \times {\vec e_i})}{ |{\vec e_z} \times {\vec e_i}| },
\ee
where we have chosen the line of the major-axis to be parallel to $z$-axis
and 
\be
{\vec e_i}=(\cos \phi_i \sin \theta_i, \sin \phi_i \sin \theta_i, \cos \theta_i), \label{unit-vector-ei}
\ee
is the unit vector pointing to the $i$-th BH (see Fig.~\ref{fig-tidal}).
For the Keplarian motion, there is a relation between the orbital angular momentum
and the eccentricity $e$ as
\be
|{\vec J}|=m_1 m_2 \sqrt{\frac{G D}{m_{\rm t}}} \sqrt{1-e^2}.
\ee
Using this formula, we obtain 
\be
1-e^2= \frac{9}{4} {\vec \zeta}^2,~~~~~ 
{\vec \zeta}=\sum_{i=1}^N \frac{x^3}{y_i^3} \frac{M_i}{m_{\rm t}}
\sin (2\theta_i ) \frac{({\vec e_z} \times {\vec e_i})}{ |{\vec e_z} \times {\vec e_i}| }, \label{eccen-e}
\ee
where $x$ is the comoving distance between BH1 and BH2 and $y_i$ is the comoving distance to the 
$i$-th BH.
Eqs.~(\ref{major-axis}) and (\ref{eccen-e}) are the main results of
this subsection. They are the major axis and
the eccentricity of the BH binary at the time of formation.
Our analysis in the next subsection is based on these formulae.

Let us now estimate the value of $N$, namely the number of the surrounding BHs that are inside the Hubble horizon
at the time of the PBH binary formation.
For simplicity, only in this paragraph we assume all the PBHs have the same mass $m_{\rm BH}$ and constitute a fraction
$f_{\rm PBH}$ of all the cold dark matter
(for instance, $f_{\rm PBH} \simeq 10^{-3}$ is required to explain the LIGO observation \citealt{Sasaki:2016jop}).
First of all, we notice that $N$ depends on the initial comoving separation of the PBHs that form a pair.
For instance, if the initial comoving separation of the BHs that form a binary is sufficiently small,
they form a binary at very early time. 
In such a case, most likely few BHs exist inside the Hubble horizon
and $N=0$ or $N=1$ will be the typical value.
Thus, what we have to estimate is the typical value of $N$ of PBH binaries that are relevant
to observations.
According to \citet{Sasaki:2016jop}, the probability
$dP$ that a given BH pair forms a binary, and then undergoes a merger at short cosmic time interval $(t,t+dt)$ is given by 
\be
dP=\frac{3}{16} {\left( \frac{t}{T} \right)}^{3/8} e {(1-e^2)}^{-(45/16)}\frac{dt}{t} de,
\ee
where $T$ is defined by
\be
T\equiv \frac{3}{170} \frac{f_{\rm PBH}^{-16/3} {(Gm_{\rm BH})}^{-5/3}}{{(1+z_{\rm eq})}^4}
{\left( \frac{8\pi}{3H_0^2 \Omega_m} \right)}^{4/3}.
\ee
For distinction between the lifetime and merger time of binaries, 
see discussion around Eq.~(\ref{life-time-PBH-binary}).
The merger probability for fixed $t$ is dominated by the binaries having eccentricity near its upper limit 
$e_{\rm upper}$ given by Eq.~(11) in \citep{Sasaki:2016jop},
\be
\label{def-eupper}
e_{\rm upper}=\begin{cases}
    \sqrt{1-{\left( \frac{t}{T} \right)}^{\frac{6}{37}}} ~~~~~~~~~~~{\rm for}~t< f_{\rm PBH}^{37/3} T\\
    \sqrt{1-f_{\rm PBH}^2 {\left( \frac{t}{f_{\rm PBH}^{37/3} T} \right)}^{\frac{2}{7}}}~~~~~{\rm for}~t\ge 
    f_{\rm PBH}^{37/3} T.
  \end{cases}
\ee
We only consider the first case $t<f_{\rm PBH}^{37/3}T$ 
which is shown to be relevant to LIGO observations \citep{Sasaki:2016jop}.
For PBH mass $m_{\rm PBH}=30~\Msun$, this condition becomes $f_{\rm PBH} \gtrsim 10^{-3}$.
Analysis in the second case is straightforward.
PBH binaries we are interested in are those that merge on the order
of the age of the Universe $t=t_0\sim 1/H_0$.
Then, when we fix the merger time and the eccentricity to $t_0$ and $e_{\rm upper}$, respectively,
the major-axis $a$ at the time of the binary formation is uniquely determined (see Eq.~(\ref{life-time-PBH-binary})).
Once the typical major-axis is determined in this way, we can convert it to
the typical redshift of the PBH binary formation by using 
Eqs.~(\ref{decoupling-redshift}) and (\ref{major-axis}), from which we can evaluate the
number of PBHs inside the Hubble horizon at that redshift, namely $N$.
The result is given by
\be
N \sim 3\times 10^{10}~{\left( \frac{t}{t_0}\right)}^{9/37}
f_{\rm PBH}^{-26/37}{\left( \frac{m_{\rm BH}}{10~\Msun} \right)}^{-22/37}.
\ee
Thus, for the typical PBH binary with $m_{\rm PBH}={\cal O}(10\,\Msun)$ which we are interested in, 
there are in general more than $\sim 3\times 10^{10}$
PBHs in the Hubble horizon at the time of the binary formation if $t\simeq t_0$.
Because of the weak dependence of the PBH number $N$ on the merger time $t$,
$N$ is much bigger than unity for merger times relevant to observations.
In what follows, we take $N \to \infty$.

One may wonder if the subsequent torque exerted on the BH binary by the surrounding BHs
changes significantly the orbital parameters from the ones given by Eqs.~(\ref{major-axis}) and (\ref{eccen-e}).
Considering the contribution only from the closest BH ($i=1$) for simplicity, 
the angular momentum that the BH pair acquires during one period $T$ of the orbital motion is given by
\be
\Delta J= \frac{3}{2} \frac{GM_1 D^2}{2D_1^3} \frac{m_1m_2}{m_{\rm t}} \sin (2\theta_1) T.
\ee
While $D$ does not increase with the scale factor because the BH pair is gravitationally bound,
the distance $D_1$ grows in proportion to the scale factor which scales as $\propto t^{1/2}$
in the radiation dominated epoch.
Then, denoting by $D_1^{(0)}$ the initial value of $D_1$ at the time of binary formation,
$D_1$ when the BH pair is in the $n$-th cycle of the orbital motion becomes $n^{1/2} D_1^{(0)}$.
The accumulated angular momentum becomes
\be
J<\Delta J \sum_{n=1}^\infty n^{-3/2} \approx 2.6~ \Delta J.
\ee
Thus, the subsequent change of the angular momentum of the BH binary after its formation
is at most a factor of $\sim 2$.
This factor is not important for our main result, and we do not consider this effect in the following analysis. On the other hand, note that if a distant third black hole with mass $M_1$ is captured on a bound orbit around the binary in a hierarchical configuration with some orbital period $T_{1}\gg T$ and eccentricity $e_{1}$, it can cause significant changes in the eccentricity of the binary due to the Lidov-Kozai effect on a timescale $t_{\rm Kozai}=[(m_{\rm t}+M_1)/M_1](1-e_{1}^2)^{3/2}T_{1}^2/T$ \citep{Naoz:2016}. However, we neglect this possibility in this paper for simplicity.

There are also other effects that have been ignored in deriving Eqs.~(\ref{major-axis}) and (\ref{eccen-e}).
They include peculiar velocity of the individual BH seeded in at the time of BH formation,
the radiation drag, the tidal interaction with the other PBHs in the matter dominated epoch,
subsequent infall of the surrounding BHs to the BH binary,
tidal force from the perturbations of non-PBH dark matter,
and baryon accretion onto the PBH binaries.
The first three effects are investigated in \citep{Ioka:1998nz} and was found to be subdominant.
Recent study by \citet{Ali-Haimoud:2017rtz} also confirms that the tidal forces from
outer PBHs do not significantly affect the late-time evolution of PBH binaries.
The subsequent infall of the surrounding BHs
is also studied in \citep{Ioka:1998nz}. 
\citet{Ioka:1998nz} assumed that the dark matter consists of a single-mass PBH population.
In this case, the surrounding BH that caused the angular momentum of the BH binary at early times
is eventually trapped by the BH binary if the outer BHs are within the mean distance of PBHs,
which can be also understood from the expression of $x_{\rm max}$ given by Eq.~(\ref{eq-xmax}).  
Since the dynamics of three-body problem is difficult to solve, such a case was not considered,
and only the opposite case where the nearest BH is more distant than the mean distance 
was included in the derivation of the merger event rate in \citep{Ioka:1998nz}.
Even under this restriction, it was found that the event rate is reduced by at most by $40\%$.
On the other hand, in the present case where PBHs constitute only a fraction $f_{\rm PBH}$ of all the cold dark matter,
the mean distance is enhanced by a factor $f_{\rm PBH}^{-1/3}$ compared with the case where PBHs provide all of the dark matter.
Thus the probability that the surrounding BHs are trapped by the BH binary in the latter case is
smaller than the former by a factor $f_{\rm PBH}$.
Because of this consideration, we make an assumption that the surrounding BHs 
are not gravitationally bound to the BH pair.
Then, the subsequent interaction by the surrounding BH
in the BH binary is not significant, and we ignore the late-time effect of the surrounding BHs in the following analysis.

The tidal force from the surrounding density perturbations of cold dark matter not in the form of PBHs, 
exists when PBHs constitute only a fraction of entire dark matter. 
This issue was addressed by \citet{Eroshenko:2016hmn} and \citet{Ali-Haimoud:2017rtz} 
who showed that the tidal effect is not significant by extrapolating the primordial perturbations on CMB scales
down to the PBH scales (see also \citealt{Hayasaki:2009ug}).
Due to the random nature of the density perturbations, they yield additional statistically independent random contribution to ${\vec \zeta}$ in Eq.(14). 
Since the power of the dark matter perturbation on small scales is not well understood,
we do not consider this effect in this paper.

Finally, baryon accretion onto PBHs was claimed to significantly affect
the PBH binaries and accelerate mergers in \citealt{Hayasaki:2009ug}.
But, recent study by \citet{Ali-Haimoud:2017rtz}, based on the simple analytic calculation,
suggests that the baryon mass accumulated on PBHs in \citealt{Hayasaki:2009ug} is likely to be an overestimation
and the baryonic effect is much weaker although it may still be significant with respect to angular momentum exchange. For simplicity we do not account for baryon accretion in this work.

\section{Distribution of the merger rate}
\label{sec-dmr}
In the previous section, we have derived the expressions for the major axis and
the eccentricity of the PBH binary in terms of the initial comoving positions and
masses of PBHs. 
They are the basic ingredients for the evaluation of the merger rate, which is the purpose
of this section.

Let us denote by ${\cal R} (m_1,m_2,t)$ a merger event density per unit cosmic time $t$
and unit comoving volume in the $m_1-m_2$ plane.
In other words,
\be
{\cal R} (m_1,m_2,t) dm_1 dm_2 dt dV,
\ee
represents the number of merger events of PBH binaries in the mass intervals $(m_1,m_1+dm_1), (m_2,m_2+dm_2)$ 
that happen during $(t,t+dt)$ and in the comoving volume $dV$.
Since the merger time $t$ can be inferred from the luminosity distance 
(depending on the cosmological parameters), and the source frame BH masses $(m_1,m_2)$
can be also estimated from the GW waveform, 
${\cal R}$ is the quantity that can be in principle determined observationally.
Our strategy to derive ${\cal R}(m_1,m_2,t)$ is described as follows. 
What we have to evaluate is the probability $P_{\rm intr} (m_1,m_2,t)dt$ 
that a given BH pair consisting of two BHs with $m_1$ and $m_2$, respectively, forms a binary, 
and then undergoes a merger during the short cosmic time interval $(t,t+dt)$. 
Once the quantity $P_{\rm intr}$ is obtained, using the PBH mass function given by Eq.~(\ref{pbh-mf})
and assuming that the masses of the two PBHs in the binary are independent, 
the merger rate density ${\cal R}$ is given by
\be
\mathcal{R}(m_1,m_2,t)=\frac{n_{\rm BH}}{2} f(m_1)f(m_2)P_{\rm intr}(m_1,m_2,t). \label{obs-P}
\ee
The major-axis and the eccentricity of the BH binary at the formation time are given by
Eqs.~(\ref{major-axis}) and (\ref{eccen-e}), respectively.
From these equations, we see that the initial semimajor axis is a function of the random variable $x$ as $a\equiv a(x)$
and the initial eccentricity is a function of the length of the random vector ${\vec \zeta}$ as $e\equiv e(\zeta)$ ,
where $\zeta=|{\vec \zeta}|$.
Denoting by $F$ the probability distribution for $x$ and $\zeta$,
the probability that the BH binary takes the values of the parameters in the range $(x,x+dx)$ and $(\zeta,\zeta+d\zeta)$
is given by
\be
F(x, \zeta) dx d\zeta. \label{F-x-zeta}
\ee
We can then convert this probability into the one expressed in terms of $a$ and $e$ as
\be
F(x(a), \zeta (e)) \frac{dx}{da} \frac{d\zeta}{de} da de. \label{p-da-de}
\ee
This gives the probability that the BH binary at the formation time has the major-axis 
and the eccentricity in the range $(a,a+da),~(e,e+de)$.

PBH binaries shrink by emitting GWs until they finally merge.
The lifetime {$\tau$} of the BH binary with parameters $(m_1,m_2,a,e)$ until it merges due to GW emission 
is given by\footnote{We assume that $e$ is typically close to 1 initially, 
which is a good approximation in the present case.} \citep{Peters:1964zz}
\be
\tau=Q {(1-e^2)}^{7/2} a^4,~~~~~Q= \frac{3}{85} \frac{1}{G^3 m_1 m_2 m_{\rm t}}. \label{life-time-PBH-binary}
\ee
Denoting by $t_{\rm dec}$ the cosmic time corresponding to $z_{\rm dec}$,
namely the time of binary formation,
we have $\tau=t-t_{\rm dec}$.
Since PBH binaries that are relevant to GW observations merge at late time $t \gg t_{\rm dec}$
($t_{\rm dec} < 4\times 10^5~{\rm yr}$),
it is a good approximation to identify $\tau$ with $t$.
Thus, in what follows, we replace $\tau$ in all of the expressions with $t$.
Under this approximation,
we can express $a$ as a function of $\{ t,~e,~m_1,~m_2\}$ as $a=a(t,e,m_1,m_2)$.
Using this relation, Eq.~(\ref{p-da-de}) becomes
\be
F(x(a), \zeta(e)) \frac{dx}{da} \frac{d\zeta}{de} \frac{\partial a}{\partial t} de dt,
\ee
where it should be understood that $a$ is replaced by $\{ t,e,m_1,m_2 \}$.
Initial eccentricity of the BH binary is not a quantity that can be measured directly 
by the GW interferometers for primordial binaries and must be integrated.
There is an upper bound $e_{\rm m}$ for the initial eccentricity for fixed $t$ because of the existence of the maximum value of the major axis 
$a_{\rm max}=x_{\rm max}/(1+z_{\rm eq})$ (see Sec.\ref{subsec:Mm}).
It is determined by the equation
\be
t=Q {(1-e_{\rm m}^2)}^{\frac{7}{2}} a_{\rm max}^4. \label{def-em}
\ee 
Notice that in the case of the monochromatic mass function 
$e_{\rm m}$ coincides with $e_{\rm upper}$ in the second case in Eq.~(\ref{def-eupper}).
Finally, the intrinsic probability distribution is given by
\be
P_{\rm intr}(m_1,m_2,t) = \int_0^{e_{\rm m}} de ~F(x(a), \zeta(e)) \frac{dx}{da} \frac{d\zeta}{de} \frac{\partial a}{\partial t}. \label{formal-Pint}
\ee

Having established the general framework to compute the merger rate density,
let us implement this methodology in practice.
It is straightforward to derive the last three factors in the integrand of Eq.~(\ref{formal-Pint}),
and they are given by
\begin{align}
\frac{dx}{da} &=\frac{1}{4} {(Aa^3)}^{-1/4},~~~
\left|\frac{d\zeta}{de} \right|=\frac{2e}{3\sqrt{1-e^2}},~~~
\\
\frac{\partial a}{\partial t}&=\frac{1}{4t} {\left( \frac{t}{Q} \right)}^{1/4}
{(1-e^2)}^{-7/8}.
\end{align}
The highly non-trivial part is the evaluation of $F (x(a),\zeta(e))$ since ${\vec \zeta}$ depends on
many random variables (in fact, infinite number of variables) in a complicated manner.
Formally, it can be written as 
\begin{align}
F (x(a),\zeta(e))=& \Theta (a_{\rm max}-a) \frac{4\pi x^2 (a)}{n_{\rm BH}^{-1}}
\nonumber\\&\times
\int \lim_{N\to \infty}
\prod_{i=1}^N \frac{dV_i}{n_{\rm BH}^{-1}} \frac{f(M_i)dM_i}{n_{\rm BH}}
\frac{\sin \theta_i d\theta_i d\phi_i}{4\pi}
\nonumber \\&\times 
\Theta (y_i-y_{i-1} ) 
e^{ -\frac{4\pi}{3} n_{\rm BH} y_N^3}
\delta \left(  \zeta-g (x,y_i,M_i,\theta_i,\phi_i ) \right), \label{formal-F}
\end{align}
where $\Theta(\cdot)$ is the Heaviside step function and $\delta (\cdot)$ is the Dirac's delta function.
Here, we have used the parametrization Eq.~(\ref{unit-vector-ei}) for ${\vec e_i}$,
and introduced the notation as $y_0=x, ~dV_i=4\pi y_i^2 dy_i$ and
\be
g (x,y_i,M_i,\theta_i,\phi_i) \equiv \bigg| \sum_{i=1}^N \frac{x^3}{y_i^3} \frac{M_i}{m_{\rm t}}
\sin (2\theta_i ) \frac{({\vec e_z} \times {\vec e_i})}{| {\vec e_z} \times {\vec e_i} |} \bigg|. \label{function-g}
\ee
The derivation of Eq.~(\ref{formal-F}) is given in the appendix \ref{app-formal-F}.

We evaluate $F(x(a),\zeta(e))$ using two approximations.
The first case is that only the nearest BH ($i=1$) is incorporated in the calculation of ${\vec \zeta}$.
This approximation was adopted in the previous studies \citep{Nakamura:1997sm,Ioka:1998nz,Sasaki:2016jop} for single-mass PBH mass functions.
In that case, all the PBHs have the same mass and the nearest BH ($i=1$) exerts the strongest torque on the BH binary.
Given that the torque by an outer BH is suppressed by the inverse cube of the distance,
the approximation of taking only the nearest BH into account is physically natural as the zero-th
order approximation\footnote{\label{f6}
The cumulative torque from all objects in a logarithmic radius bin of width $\Delta \ln y$ 
(e.g. here we may set $\Delta \ln y \sim \Delta y / y \sim n_{\rm BH}^{-1/3} / y$) follows from the central limit theorem
and is described by a normal distribution with zero mean and root-mean-square that corresponds to $\Delta N^{1/2}g_{1,\rm RMS}$, 
where $\Delta N$ is the number of objects in that logarithmic radius bin and 
$g_{1,\rm RMS} = 2^{-1/2} (x/y)^{-3} (M_{\rm RMS}/m_{\rm t})\sin(2\theta)_{\rm RMS}$. 
This may be estimated roughly as $\Delta N = 4\pi n_{\rm BH} y^3 \Delta \ln y$. 
Therefore, the relative cumulative contribution of distant objects to the torque scales with $y^{-3/2}$, 
and so the smallest $y$ dominates the integral where the number of objects is $\sim 1$.}.

On the other hand, if the mass function is multimass,
a massive outer BH may exert a stronger torque than a low-mass inner one.
The wider the mass function, the more likely it is that this possibility may arise.
To take into account the effect of outer perturbers, in our second estimate 
we consider a flat mass function up to a certain BH mass $m_{\rm max}$ and 
include the outer BHs to evaluate the torque.

In what follows, we evaluate $F (x(a),\zeta(e))$ and the intrinsic probability distribution for these two cases, separately.

\subsection{Case 1: torque only from the nearest BH}
In this subsection, we make an approximation that the torque is exerted only by the nearest BH.
Accordingly, the function $g$ defined by Eq.~(\ref{function-g}) becomes
\be
g= \frac{x^3}{y_1^3} \frac{M_1}{m_{\rm t}} \sin (2\theta_1). \label{only-nearest-g}
\ee
Even after this simplification, 
it is hard to evaluate the integral (\ref{formal-Pint}) analytically.
For an analytic estimate, we carry out the
calculation for an arbitrary but fixed value $\beta=\sin (2\theta_1)$.
Our result is insensitive to the value of $\beta$ as long as it is not extremely close to zero.
Since the probability of realizing $\beta \ll 1$ is suppressed (see discussion after Eq.~(\ref{case1-expression-F})
for the estimation of this probability), 
we think that this simplification
does not lose the essential feature of the merger-rate density.
The integral over $y_1$ simplifies to
\begin{align}
F(x(a),\zeta (e))=&\Theta (a_{\rm max}-a) \frac{ 12\pi^2 n_{\rm BH}}{1-e^2}\beta {\left( \frac{a}{A} \right)}^{5/4} 
\nonumber \\&\times
\int dM_1 f(M_1) \frac{M_1}{m_{\rm t}} \exp \left( -\frac{2\pi n_{\rm BH} M_1}{\sqrt{1-e^2}m_{\rm t}} {\left( \frac{a}{A} \right)}^{3/4}\beta \right) 
\nonumber \\&\times
\Theta \left( \frac{M_1}{m_{\rm t}}\beta-\frac{2}{3} \sqrt{1-e^2} \right). \label{case1-expression-F}
\end{align}
The PBH binaries at the time of their formation are highly eccentric ($e \approx 1$).
Since the PBH mass function is implicitly assumed to be narrow in the present case,
$M_1$ does not differ from $m_{\rm t}$ significantly, and the argument of the last Heaviside function is positive
unless $\beta$ is smaller than $\frac{2}{3} \frac{m_{\rm t}}{M_1} \sqrt{1-e^2}$.
Now, let us estimate the probability that $\beta$ becomes smaller than the critical value $\beta_c$
for which the argument of the Heaviside function becomes zero.
To this end, we again consider the monochromatic mass function and use the eccentricity given by 
the first case of Eq.~(\ref{def-eupper}). Then, $\beta_c$ becomes
\be
\beta_c \simeq 0.01 \times f_{\rm PBH}^{16/37} {\left( \frac{t}{t_0} \right)}^{3/37}
{\left( \frac{m_{\rm BH}}{10\,\Msun} \right)}^{5/37}. \label{beta_c}
\ee
For $\beta_c \ll 1$, the probability that $\beta$ happens to be smaller than $\beta_c$
is approximately given by
\be
P(\beta < \beta_c )\approx \frac{\beta_c^2}{16} \simeq 6\times 10^{-6},
\ee
for the fiducial values used in Eq.~(\ref{beta_c}).
This probability is much smaller than unity, 
and we replace the last Heaviside function by $1$ in the following analysis.
Then, the intrinsic probability distribution (\ref{formal-Pint}) becomes
\begin{align}
P_{\rm intr}(m_1,m_2,t) &=\frac{1}{8t} \int dM_1~\frac{1}{\beta}
\frac{m_{\rm t}}{M_1} K^2 \frac{f(M_1)}{n_{\rm BH}}
\nonumber\\&\times
\int_0^{e_{\rm m}} de~e {(1-e^2)}^{-\frac{45}{16}} \exp \bigg[ -K {(1-e^2)}^{-\frac{37}{32}} \bigg], \label{case1-Pint}
\end{align}
where we have introduced a dimensionless parameter $K$ by
\be
K \equiv 2\pi n_{\rm BH} \frac{M_1}{m_{\rm t}} A^{-\frac{3}{4}} {\left( \frac{t}{Q} \right)}^{\frac{3}{16}} \beta.
\label{definition-of-K}
\ee
This is a small parameter. For instance, for a single-mass PBH mass function
with mass $m_{\rm BH}$ and the Hubble time $t=1/H_0$, we have
\begin{align}
K &= {\left( \frac{170}{3} \right)}^{\frac{3}{16}} {\left( \frac{3}{\pi} \right)}^{\frac{1}{4}}
{(1+z_{\rm eq})}^{\frac{3}{4}} \pi \Omega_m^{\frac{1}{4}} f_{\rm PBH} {(G m_{\rm BH} H_0 )}^{\frac{5}{16}}\beta
\nonumber\\&\sim 
3 \times 10^{-4}~f_{\rm PBH} \beta {\left( \frac{m_{\rm BH}}{10 \Msun} \right)}^{\frac{5}{16}},
\label{estimate-k}
\end{align}
where $f_{\rm PBH}$ is the mass fraction of the PBHs to the entire cold dark matter

The integration over $e$ can be expressed in terms of the incomplete gamma function.
Then, Eq~(\ref{case1-Pint}) becomes
\begin{align}
P_{\rm intr}(m_1,m_2,t) =&\frac{2}{37 t} \int dM_1~\frac{1}{\beta} 
\frac{m_{\rm t}}{M_1} \frac{f(M_1)}{n_{\rm BH}} K^{\frac{16}{37}}
\nonumber\\&\times
\bigg[ G(K)-G \left( \frac{M_1}{m_c} \right) \bigg], \label{case1-Pint2}
\end{align}
where $m_c$ and $G(x)$ are defined by
\begin{align}
&m_c \equiv
\frac{m_{\rm t}}{2 \pi n_{\rm BH}}\frac{1}{\beta} {\left( \frac{t}{Q} \right)}^{1/7} {(1+z_{\rm eq})}^{4/7}
{\left( \frac{\rho_{c,0} \Omega_m}{m_{\rm t}} \right)}^{\frac{25}{21}}, \label{definition-of-mc} \\
&G(x)=\Gamma \left( \frac{58}{37}, x \right). \label{definition-of-G(x)}
\end{align}
For the monochromatic mass function, $m_c$ is given by
\be
m_c \sim 7\times 10^{-4} \Msun ~{(f_{\rm PBH} \beta)}^{-1}
{\left( \frac{m_{BH}}{10 \Msun} \right)}^{\frac{26}{21}}.
\ee
Eq.~(\ref{case1-Pint2}) for arbitrary $f(M)$ mass function is the final expression of the intrinsic merger probability distribution in the present case.

\subsection{Case 2: torque from the outer BHs}
\label{case2}
Let us next consider the case in which the PBH mass function is flat from 
$m_{\rm min}=\epsilon m_{\rm max}$ to $m_{\rm max}$
and vanishes outside of it. 
As mentioned earlier, we implicitly assume that $\epsilon \gtrsim 0.1$.
Then the PBH mass function is given by
\be
f(m)=\frac{1}{m_{\rm max} (1-\epsilon)}
\Theta (m_{\rm max}-m ) \Theta (m-m_{\rm min} ). \label{flatmf}
\ee 
We include not only the nearest BH but also outer BHs.

It is extremely difficult to perform the integration of Eq.~(\ref{formal-F}) 
analytically \footnote{Analytic expression of the probability distribution 
for the eccentricity was derived for the monochromatic mass function in \citet{Ali-Haimoud:2017rtz}.}.
However, we can estimate the approximate behavior of $F(x,\zeta)$ in the domain $n_{\rm BH} x^3 \ll 1$
where the PBH binaries with lifetime comparable to the age of the Universe form 
\footnote{PBH binaries with $n_{\rm BH}x^3 \sim 1$ have larger semimajor axis
and more circular orbit than those with $n_{\rm BH}x^3 \ll 1$.
These two factors make the lifetime of the binaries much longer than the age of the Universe.}.
To this end, let us first write $F(x,\zeta)$ as
\be
F(x,\zeta)=\frac{4\pi x^2}{n_{\rm BH}^{-1}} e^{-\frac{4\pi}{3}n_{\rm BH} x^3} P(x,\zeta),
\ee
where $P(x,\zeta_0)d\zeta$ is a probability that $\zeta$ takes value in the interval $(\zeta_0,\zeta_0+d\zeta)$
for given $x$.  
For later convenience, let us define ${\tilde \zeta}$ by $(m_{\rm t}/m_{\rm max}) \zeta$. Thus, we have
\be
F(x,\zeta) \approx 4\pi n_{\rm BH}x^2 {\tilde P}(x,{\tilde \zeta}) \frac{m_{\rm t}}{m_{\rm max}},
\ee
where ${\tilde P}(x,{\tilde \zeta}_0)d{\tilde \zeta}$ is the probability that ${\tilde \zeta}$ takes a value 
in the interval $({\tilde \zeta}_0, {\tilde \zeta}_0+d{\tilde \zeta})$ for given $x$. 
Looking at the definition of ${\vec \zeta}$,
we expect that the typical value of ${\tilde \zeta}$ for given $x$ is around $n_{\rm BH}x^3$
since $y_i$ ($i={\cal O}(1)$) is typically about $n_{\rm BH}^{-1/3}$ and the contribution from
$y_i$ with higher $i$ is suppressed (see footnote \ref{f6}).
Noting that $y_i > x$, the case in which $\zeta \ll n_{\rm BH}x^3$ is realized by
either if $y_1 \gg n_{\rm BH}^{-1/3}$ or if accidental cancellation takes place among terms with different $i$.
Since the former is suppressed exponentially as $\sim e^{-\frac{4\pi}{3}n_{\rm BH}y_1^3}$,
the latter, which is stochastic, dominates.
Recalling that ${\vec \zeta}$ is essentially a two-dimensional vector,
the probability that ${\tilde \zeta}$ is in the thin ring $({\tilde \zeta},{\tilde \zeta}+d{\tilde \zeta})$ by the random choice
is proportional to the ring area, namely ${\tilde \zeta} d{\tilde \zeta}$. Thus, we expect
\be
{\tilde P}(x,{\tilde \zeta}) \propto {\tilde \zeta}, \label{tilde-P-1}
\ee
for ${\tilde \zeta} \ll n_{\rm BH}x^3$.
On the other hand, the case ${\tilde \zeta} \gg n_{\rm BH}x^3$ is realized mainly when $y_1$ is accidentally
much smaller than the typical value $n_{\rm BH}^{-1/3}$.
The probability of such a situation is controlled by the volume element $y_1^2 dy_1$, and
the relation ${\tilde \zeta} \propto y_1^{-3}$ leads to $y_1^2 dy_1 \propto {\tilde \zeta}^{-2} d{\tilde \zeta}$.
Thus, we expect
\be
{\tilde P}(x,{\tilde \zeta}) \propto {\tilde \zeta}^{-2}, \label{tilde-P-2}
\ee
for ${\tilde \zeta} \gg n_{\rm BH}x^3$.
From the definition of ${\vec \zeta}$ given by Eq.~(\ref{eccen-e}),
we have
\be
\sigma^2 \equiv \langle {\vec \zeta}^2 \rangle= \frac{32 \pi}{135} {\left( \frac{m_{\rm max}}{m_{\rm t}} \right)}^2 
n_{\rm BH} x^3 (1+\epsilon+\epsilon^2). \label{flat-variance}
\ee
The derivation of this result is given in appendix \ref{app-flat-variance}.
One simple function that interpolates Eqs.~(\ref{tilde-P-1}) and (\ref{tilde-P-2}) is given by
\be
{\tilde P} (x,{\tilde \zeta}) = \frac{3\sqrt{3}}{2\pi} \xi^{1/3} {\tilde \sigma}^2 
\frac{\tilde \zeta}{{\tilde \zeta}^3+\xi {\tilde \sigma}^6}, \label{P-ansatz}
\ee
where ${\tilde \sigma}= (m_{\rm t}/m_{\rm max}) \sigma$, $\xi={\cal O}(1)$ is a fitting parameter,
and the normalization condition is imposed.

In order to check the validity of the approximation (\ref{P-ansatz}),
we evaluate ${\tilde P}(x,{\tilde \zeta})$ numerically by the Monte Carlo method.
For this purpose, we first fix $N$ and $x$. 
Then, we randomly generate a set of random variables
$\{ M_i, y_i, \theta_i, \phi_i \}$ 
and compute ${\tilde \zeta}$.
By repeating this process many times, we obtain the distribution of ${\tilde \zeta}$
for a given $N$ and $x$ up to the statistical uncertainty.

\begin{figure*}
\begin{center}
\begin{tabular}{cc}
\begin{minipage}[t]{0.5\hsize}
\includegraphics[width=8cm]{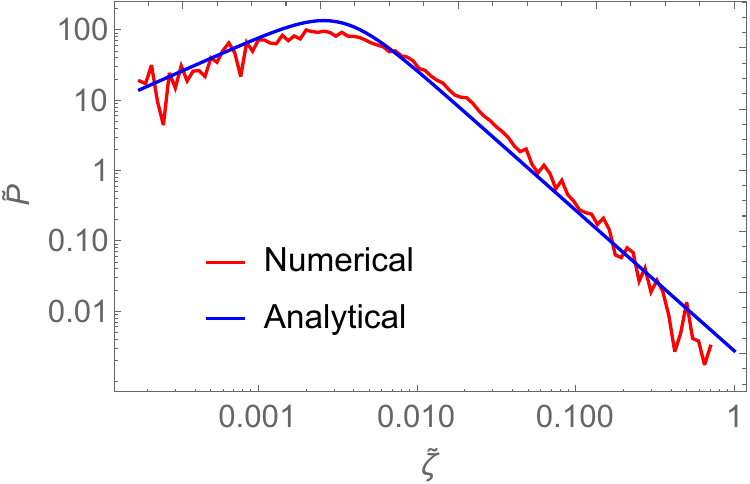}
\end{minipage}
\begin{minipage}[t]{0.5\hsize}
\includegraphics[width=8cm]{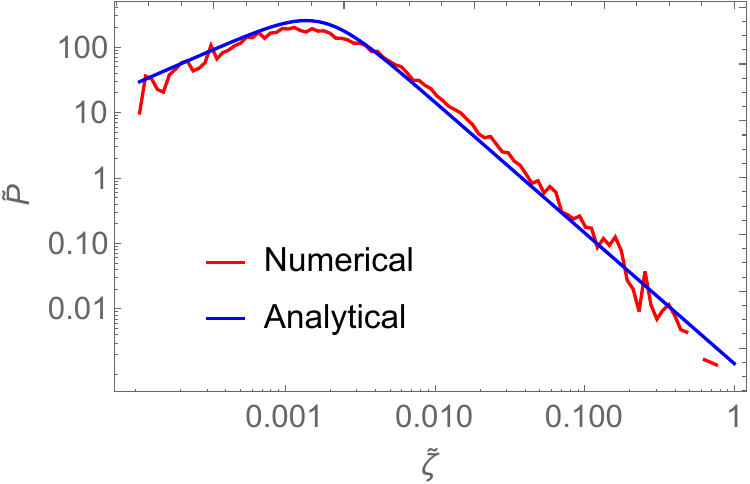}
\end{minipage}
\\
\begin{minipage}[c]{0.5\hsize}
\includegraphics[width=8cm]{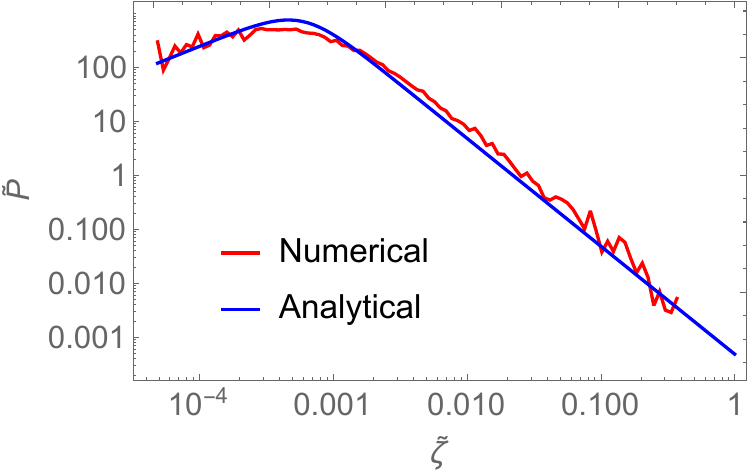}
\end{minipage}
\begin{minipage}[c]{0.5\hsize}
\includegraphics[width=8cm]{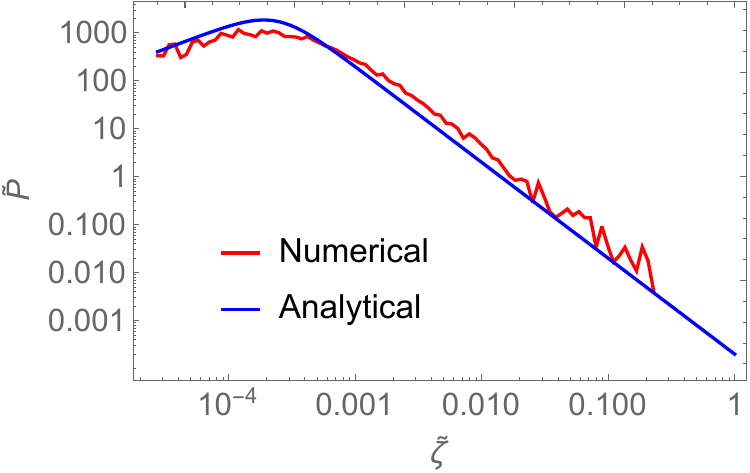}
\end{minipage}
\end{tabular}
\end{center}
\caption{Red curves represent the probability distribution ${\tilde P}(x,{\tilde \zeta})$ of ten thousand Monte Carlo realizations of the dimensionless torque parameter ${\tilde \zeta}$
for four different values of $\frac{4}{3}\pi n_{\rm BH} x^3 =10^{-2}$ (top-left panel),
$5\times 10^{-3}$ (top-right panel), $2\times 10^{-3}$ (bottom-left panel), and $10^{-3}$ (bottom-right panel)
with $N=5$ perturbing BHs
for a flat PBH mass function (\ref{flatmf}) 
with $\epsilon=0.1$.
Here $n_{\rm BH}$ and $x$ is the comoving PBH number density and
initial comoving distance between BHs that form binary, respectively.
Blue curves represent the probability distribution given by Eq.~(\ref{P-ansatz}) with $\xi=5.5$.
}
\label{MC}
\end{figure*}

Figure~\ref{MC} shows the distribution of ten thousand realizations of ${\tilde \zeta}$ 
for $N=5$ for $\epsilon=0.1,~\frac{4\pi}{3} n_{\rm BH} x^3=(10^{-2},5\times 10^{-3},2\times 10^{-3}, 10^{-3})$.
The red curve represents the distribution obtained by the Monte Carlo calculations,
and blue one represents the analytic approximation (\ref{P-ansatz}) with $\xi =5.5$.
We find that this simple ansatz of ${\tilde P}(x,{\tilde \zeta})$ 
fairly recovers
the numerically obtained probability distribution.
Although we consider the flat mass function, we expect that the ansatz should work qualitatively
for other mass functions since the asymptotic behaviors (\ref{tilde-P-1}) and (\ref{tilde-P-2})
are determined independently of the mass function. 
In what follows, we adopt Eq.~(\ref{P-ansatz}). Then, $F(x,\zeta)$ becomes
\be
F(x,\zeta)=6\sqrt{3} \xi^{1/3} n_{\rm BH} {\tilde \sigma}^2 \zeta x^2 {\left( \frac{m_{\rm t}}{m_{\rm max}} \right)}^2 
{\bigg[ {\left( \frac{m_{\rm t}}{m_{\rm max}} \right)}^3 \zeta^3+\xi {\tilde \sigma}^6 \bigg]}^{-1}. \label{flat-F-ansatz}
\ee

Substituting $F(x,\zeta)$ given by Eq.~(\ref{flat-F-ansatz}) into Eq.~(\ref{formal-Pint}), after some algebra, we obtain 
\be
P_{\rm intr} =\frac{135 \sqrt{3}}{256\pi t} 
\frac{1}{\xi^{1/3} (1+\epsilon+\epsilon^2)}
\nu^{\frac{16}{37}} \frac{m_{\rm t}}{m_{\rm max}}
\int_{w_{\rm m}}^\infty \frac{w^{\frac{21}{32}}}{w^{\frac{111}{32}}+1} dw. \label{case2-Rintr}
\ee
Here we have defined a dimensionless quantity $\nu$ by
\be
\nu =\frac{16\pi}{45} \xi^{1/3} (1+\epsilon+\epsilon^2) n_{\rm BH} 
\frac{m_{\rm max}}{m_{\rm t}} A^{-3/4} {\left( \frac{t}{Q} \right)}^{\frac{3}{16}}, \label{definition-nu}
\ee
and we have changed the integration variable as $w=\nu^{-32/37} (1-e^2)$,
and $w_{\rm m}=\nu^{-32/37} (1-e_{\rm m}^2)$.
Using a relation $n_{\rm BH}=2\rho_{\rm BH}/(m_{\rm max} (1+\epsilon))$,
which is valid for a flat mass function, 
we have
\begin{align}
w_{\rm m} =& {\left( \frac{32\pi}{45} \gamma^{1/3} \frac{1+\epsilon+\epsilon^2}{1+\epsilon} \right)}^{-\frac{32}{37}}
{(1+z_{\rm eq})}^{\frac{128}{259}} f_{\rm PBH}^{-\frac{32}{37}}
{\left( \frac{\rho_{c,0}\Omega_m}{m_{\rm t}} \right)}^{\frac{128}{777}}
\nonumber\\&\times
{\left( \frac{G^3 m_1m_2m_{\rm t}}{3}t \right)}^{\frac{32}{259}}. \label{definition-zm}
\end{align}
To estimate typical magnitude of $w_{\rm m}$,
for equal mass binary ($m_1=m_2=m_{\rm BH}$), $w_{\rm m}$ is given by
\be
w_{\rm m} \approx 2\times 10^{-4} ~f_{\rm PBH}^{-\frac{32}{37}} {\left( \frac{m_{\rm BH}}{\Msun} \right)}^{\frac{160}{777}}.
\ee
This shows that $w_{\rm m}$ can be bigger or smaller than unity
within the range of the feasible values of $f_{\rm PBH}$ and $m_{\rm BH}$.
Although the integration over $w$ in Eq.~(\ref{case2-Rintr}) can be expressed in terms of the hypergeometric function,
we do not write it explicitly here since it gives no useful information.
Thus, Eq.~(\ref{case2-Rintr}) is the final expression of the intrinsic merger rate and the main result of this subsection.

\section{Hidden universality in the merger rate density}
\label{hidden}
In the previous section, we have derived the analytic expression of $P_{\rm intr}$ 
in the $m_1-m_2$ plane for the two different limiting cases 
corresponding to the different approximations.
According to Eq.~(\ref{obs-P}), the observable merger rate density is not $P_{\rm intr}$, but
$P_{\rm intr}$ weighted by the PBH mass function.
The observable merger event density is highly dependent on the PBH
mass function, and it appears at first glance that no definite prediction can be extracted 
for the PBH scenario without choosing the specific mass function.
Contrary to this naive guess, there is a unique feature expressed as a mathematical 
relation for the differentiated merger rate density specific to the PBH scenario as we will show below.
Such a relation could be quite useful as a powerful method for testing 
the PBH scenario when the sufficient number of merger events have been accumulated.

Let us first consider the case where $P_{\rm intr}$ is given by
Eq.~(\ref{case1-Pint}). This expression of $P_{\rm intr}$ still contains 
the integration over the PBH mass 
nearest to the BH binary.
Although this integration cannot be done explicitly without choosing the specific PBH mass function,  
carrying out the explicit integration is not needed for our present purpose.  
The function $G(x)$ appearing in the integrand is monotonically decreasing and 
its asymptotic behavior is given as
\be
G(x)=
\begin{cases}
\frac{21}{37} \Gamma \left( \frac{21}{37} \right)
-\frac{37}{58} x^{\frac{58}{37}}+{\cal O} \left( x^{\frac{95}{37}} \right),~~~~~(x \ll 1)\\
x^{\frac{21}{37}}e^{-x} \left( 1+{\cal O}(x^{-1}) \right).~~~~~(x \gg 1).
\end{cases}
\ee
Using this formula and noting that $K$,
which is much smaller than unity according to Eq.~(\ref{estimate-k}), 
is always less than $M_1/m_c$, 
we find that the integrand of Eq.~(\ref{case1-Pint}) becomes
\begin{align}
&\frac{m_{\rm t}}{M_1} \frac{f(M_1)}{n_{\rm BH}} K^{\frac{16}{37}}
\bigg[ G(K)-G \left( \frac{M_1}{m_c} \right) \bigg]
\nonumber\\
&\qquad =
\begin{dcases}
\frac{37}{58} \frac{m_{\rm t}}{M_1} \frac{f(M_1)}{n_{\rm BH}} K^{\frac{16}{37}}
{\left( \frac{M_1}{m_c} \right)}^{\frac{58}{37}}, ~~~\frac{M_1}{m_c}<1 \\
\frac{21}{37} \Gamma \left( \frac{21}{37} \right) 
\frac{m_{\rm t}}{M_1} \frac{f(M_1)}{n_{\rm BH}} K^{\frac{16}{37}}, ~~~\frac{M_1}{m_c}>1  \\
\end{dcases}
\end{align}
A crucial consequence of these approximate expression is that the
integrand has a simple scaling property with $m_1$ and $m_2$.
Using the scalings,
\be
K \propto m_{\rm t}^{-\frac{1}{16}} {(m_1m_2)}^{\frac{3}{16}},~~~~~
m_c \propto m_{\rm t}^{-\frac{1}{21}} {(m_1m_2)}^{\frac{1}{7}},
\ee
we find that the above integrand scales as
\begin{align}
\frac{m_{\rm t}}{M_1} \frac{f(M_1)}{n_{\rm BH}} K^{\frac{16}{37}}
&\bigg[ G(K)-G \left( \frac{M_1}{m_c} \right) \bigg]
\nonumber\\ &\propto 
\begin{cases}
{m_{\rm t}}^{\frac{22}{21}} {(m_1 m_2)}^{-\frac{1}{7}},~~~~~\frac{M_1}{m_c}<1\\
{m_{\rm t}}^{\frac{36}{37}} {(m_1 m_2)}^{\frac{3}{37}},~~~~~\frac{M_1}{m_c}>1.
\end{cases}
\end{align}
Because of this factorization, the same scaling for $m_1 m_2$ and $m_{\rm t}$ remains for $P_{\rm intr}$.
Assuming one of the branches ($M_1 < m_c$ or $M_1 > m_c$) dominates the integral,
$P_{\rm intr}$ scales as
\be
P_{\rm intr} (m_1, m_2,t) \propto 
\begin{cases}
{m_{\rm t}}^{\frac{22}{21}} {(m_1 m_2)}^{-\frac{1}{7}},~~~~~(M_1 < m_c~{\rm dominates})\\
{m_{\rm t}}^{\frac{36}{37}} {(m_1 m_2)}^{\frac{3}{37}},~~~~~(M_1 > m_c~{\rm dominates}).
\end{cases}
\ee
Then, the observable merger rate density $\mathcal{R}$ per unit time and unit volume
defined by Eq.~(\ref{obs-P}) can be written as
\be
\mathcal{R}(m_1,m_2,t)=
\begin{cases}
C_A {m_{\rm t}}^{\frac{22}{21}} h_A(m_1) h_A(m_2),~~~(M_1 <m_c~{\rm dominates}) \\
C_B {m_{\rm t}}^{\frac{36}{37}} h_B(m_1) h_B(m_2),~~~(M_1 > m_c~{\rm dominates})
\end{cases}
\label{event-rate}
\ee
where $h_A(m) \equiv m^{-\frac{1}{7}} f(m),~h_B(m) \equiv m^{\frac{3}{37}} f(m)$ and 
$C_A,~C_B$ are quantities that are independent of $m_1$ and $m_2$, but contain information of $f(m)$.
An interesting point of Eq.~(\ref{event-rate}) is that the dependence of the 
merger rate density on the total mass $m_{\rm t}$ is independent of the model-dependent
functions $h_A(m)$ or $h_B(m)$ (namely, mass function) and is completely determined as 
$\propto {m_{\rm t}}^{36/37}$ for the former case and $\propto {m_{\rm t}}^{22/21}$ for 
the latter case.
The mass function enters the game only through the total normalization constant 
(represented as $C_A$ and $C_B$)
and the factorizable part $h_A(m_1) h_A(m_2)$ or $h_B(m_1) h_B(m_2)$.
Thus, by focusing on the total mass part of merger rate density and picking it up,
we can provide a definite prediction for the merger rate density which
is insensitive to the shape and amplitude of the PBH mass function.
Indeed, we can pick up the total mass part by taking the logarithm of $\mathcal{R}$ and then
differentiating it by $m_1$ and $m_2$, namely
\begin{align}
\alpha (m_1,m_2,t) &\equiv -{m_{\rm t}}^2 \frac{\partial^2}{\partial m_1 \partial m_2} \ln \mathcal{R}(m_1,m_2,t)
\nonumber \\ &=
\begin{cases}
\frac{36}{37},~~~~~(M_1 < m_c~{\rm dominates}) \\[1.5ex]
\frac{22}{21},~~~~~(M_1 > m_c~{\rm dominates})
\end{cases}
\label{consistency}
\end{align}
for any $(m_1,~m_2)$.
As discussed at the beginning of Sec.~\ref{sec-dmr},
the merger rate density ${\cal R}$ can be determined in principle by observations
if a sufficient number of BH merger events are detected and the potential detection bias can be
appropriately eliminated.
Thus, the quantity $\alpha$ on the left-hand side can be also determined observationally.
In this sense, the left hand side can be determined by observations.
Our PBH merger scenario predicts that this quantity is equal to $36/37$ for the upper case
and $22/21$ for the lower case.
In reality, what is realized lies between the above two cases,
and the left hand side of Eq.~(\ref{consistency}) may take a value 
between the two values corresponding to the upper case and the lower case respectively.
Given that the numerical values on the right hand side for both cases are close to $1$ (within less than $5\%$),
the left hand side of Eq.~(\ref{consistency}) in the mixture case would be also 
close to $1$.
Taking into account this possibility, we conclude that under the assumption of
the uniform spatial distribution of PBHs the merger rate density
satisfies the following relation
\be
\frac{36}{37} \le 
\alpha (m_1,m_2,t)
\le \frac{22}{21}. \label{case1-consistency}
\ee
This relation is robust in the sense that it is independent of the underlying mass function.

Similar conclusion can be drawn to the second case where 
$P_{\rm intr}$ is given by Eq.~(\ref{case2-Rintr}).
In this case, the observable merger rate density (Equation~\ref{obs-P}) is given by
\be
\mathcal{R} =\frac{135 \sqrt{3}}{512\pi t} 
\frac{\nu^{\frac{16}{37}}}{\xi^{1/3} (1+\epsilon+\epsilon^2) {(1-\epsilon)}^2 }
\frac{n_{\rm BH}}{m_{\rm max}^2}
\frac{m_{\rm t}}{m_{\rm max}}
\int_{w_{\rm m}}^\infty \frac{w^{\frac{21}{32}}}{w^{\frac{111}{32}}+1} dw. \label{case2-R}
\ee
As we have done in the case 1, let us evaluate the integral for two limiting cases
($w_{\rm m} \ll 1$ and $w_{\rm m} \gg 1$), separately.

First, when $w_{\rm m} \ll 1$, we can extend the lower limit of the integral to $0$.
As a result, we obtain
\be
\mathcal{R} =\frac{C_1}{t} 
\nu^{\frac{16}{37}} {\left( \frac{n_{\rm BH}}{m_{\rm max}} \right)}^2
\frac{m_{\rm t}}{m_{\rm max}},
\ee
where $C_1$ is a constant of order unity.
Using the scaling for $\nu$ as (see Eq.~(\ref{definition-nu}))
\be
\nu \propto m_{\rm t}^{-\frac{1}{16}} {(m_1m_2)}^{\frac{3}{16}},
\ee
$\mathcal{R}$ can be written as
\be
\mathcal{R}(m_1,m_2,t)={\tilde C_1} m_{\rm t}^{\frac{36}{37}} h_1(m_1) h_1(m_2),
\ee
where $h_1(m) \equiv m^{\frac{3}{37}} f(m)$ and 
${\tilde C_1}$ is a quantity that is independent of $m_1,~m_2$, but contains information of $f(m)$.
As with the above discussion for the case 1, $\mathcal{R}$ has a unique dependence on $m_{\rm t}$.
This dependence can be again extracted by considering the quantity $\alpha$ as
\be
\alpha (m_1,m_2,t)=\frac{36}{37}. 
\label{case2-consistency}
\ee
This value precisely coincides with the lower end of Eq.~(\ref{case1-consistency}).

Let us next investigate the case $w_{\rm m} \gg 1$. In this case, we obtain
\be
\mathcal{R} \approx \frac{C_2}{t} 
\nu^{\frac{16}{37}}
{\left( \frac{n_{\rm BH}}{m_{\rm max}} \right)}^2
\frac{m_{\rm t}}{m_{\rm max}} w_{\rm m}^{-\frac{29}{16}},
\ee
where $C_2$ is a constant of order unity.
Using the scaling for $w_{\rm m}$ as (see Eq.~(\ref{definition-zm}))
\be
w_{\rm m} \propto m_{\rm t}^{-\frac{32}{777}} {(m_1m_2)}^{\frac{32}{259}},
\ee
as well as that for $\nu$, we find
\be
\mathcal{R}(m_1,m_2,t)={\tilde C_2} m_{\rm t}^{\frac{22}{21}} h_2(m_1) h_2(m_2),
\ee
where $h_2(m) \equiv m^{-\frac{1}{7}} f(m)$ and 
${\tilde C_2}$ is a quantity that is independent of $m_1,~m_2$, but contains information of $f(m)$.
Then, we find
\be
\alpha (m_1,m_2,t)=\frac{22}{21}. 
\label{case2-consistency2}
\ee
This value precisely coincides with the upper end of Eq.~(\ref{case1-consistency}).
Thus, the range of $\alpha$ in the present case is also given by Eq.~(\ref{case1-consistency}).

To summarize, our study demonstrates that $0.97 \lesssim \alpha \lesssim 1.05$ holds in the considered PBH scenario in which PBHs form binaries in the early universe.
The uncertainty in $\alpha$ is small enough to distinguish the PBH
scenario from different scenarios for explaining the origin of the merging BH binaries once a sufficiently large number of merger events are measured.
For instance, \citet{Bird:2016dcv} considered the formation of PBH binaries due to close encounters
in dark matter halos at low redshifts. This PBH scenario gives a different merger rate density, i.e. \citep{Raidal:2017mfl}
\be
{\cal R}(m_1,m_2,t)=C m_1^{\frac{2}{7}} f(m_1) \,m_2^{\frac{2}{7}} f(m_2) \,m_{\rm t}^{\frac{10}{7}},
\ee
where $C$ is a quantity independent of $m_1$ and $m_2$. For this process, Equation~(\ref{alphadef}) gives
\be
\alpha=\frac{10}{7} \approx 1.43.
\ee
Thus, this scenario predicts a unique and different value from the one studied in this paper. 
\citet{Gondanetal2017} has recently extended this analysis to systems in collisional equilibrium where mass segregation takes places such as in galactic nuclei. In this case $\alpha$ is a unique function of the total binary mass.
Another example is the astrophysical scenario in which
the BH binaries form and evolve due to dynamical encounters in dense stellar environments.
In this scenario, \citet{OLeary:2016ayz} found that approximately $P_{\rm intr}=\mathcal{R}(m_1,m_2)/[f(m_1)f(m_2)]\propto m_{\rm t}^{4}$.
In this case the higher mass mergers are much more probable mainly due to the mass dependence of binary formation during chance triple encounters, exchange interactions, mass segregation and dynamical hardening effects. 
If the intrinsic merger probability does not depend on the symmetric mass ratio $\eta=m_1m_2/(m_1+m_2)^2$, 
then we get $\alpha=4$ for this process.
Clearly, a $\alpha\sim 4$ value is largely outside of the region obtained for both PBH scenarios mentioned above.
When a sufficient number of mergers accumulates to determine $\alpha$,
it may be possible to exclude several formation scenarios and pin down the most likely scenario. 

In order to crudely estimate the necessary sample size to measure $\alpha$ from future GW detections, 
we generate a mock Monte Carlo sample of BHs drawn from a fiducial flat mass function between a range of masses 5 and $30\,\Msun$, 
and generate a random merger sample by randomly drawing objects with probability proportional to $(m_1+m_2)^\alpha$. For this order-of-magnitude estimate we neglect the measurement error of mass, since the mass measurement accuracy is expected to be much smaller than the range of BH masses, i.e. $\Delta m_{1,2}/m_{1,2}\sim 25\%$ for half of the sources for the design sensitivity of second generation GW instruments including Advanced LIGO, Advanced VIRGO, and KAGRA \citep{Ghoshetal2016}.
\footnote{If heavy BHs exist with mass $30\,\Msun<m_{1,2}<50\,\Msun$, the median mass measurement errors are expected to be of order $40\%$ \citep{Vitaleetal2017}.} We generate a 2D histogram of events and fit the value of $\alpha$. Repeating this analysis 1000 times for fixed fiducial $\alpha$ gives an approximate posterior distribution function of the measured $\alpha$.
This analysis shows that a sample of 100 events is necessary to measure $\alpha$ to 
integer accuracy and 1000 events would allow to measure it with an error of 0.15 if the fiducial value of $\alpha$ is between 1 and 3. 
The current rate estimates predict $\mathcal{R} = 12$--$240\,{\rm Gpc}^{-3}{\rm yr}^{-1}$. 
Assuming a maximum detection distance of $z=0.5$ for the design sensitivity of second generation instruments, 
a sample of $\sim 100$ events (1000 events) will accumulate in between 6 and 120 days (60 days and 3.3 years).

\section{Summary}
There is a growing interest in the possibility that the merging BHs detected by LIGO are primordial. 
Previous study \citep{Sasaki:2016jop} showed that the BH binary merger event rate estimated by LIGO
can be explained by the PBHs which constitute only a tiny fraction of the entire dark matter.
While the estimated masses of the individual BHs show some spread $10 \sim 30~\Msun$,
it was assumed in the previous study that all the PBHs have the same mass of $30~\Msun$.
Although this is a reasonable approximation when only the first event for which
masses of two BHs in the binary are almost the same is observationally known,
it hugely compresses the valuable information about 
the event rate distribution in the BH mass plane.

In this paper, we extended the formalism to compute the merger event rate to
the case where the PBH mass function is not monochromatic.
Our basic assumption on the mass function made throughout this paper is that
it is not widely extended over many orders of magnitude in the BH mass range but
is confined to the mass range $\sim 10~\Msun$. 
The derived formula (\ref{formal-Pint}) contains multiple integrations over many random variables
(Eq.~(\ref{formal-F})) and 
is complicated enough to defeat the exact analytic computation.
Based on the physical expectation that among remote BHs, the closest one gives the largest torque
on average, we evaluated the simplified version of Eq.~(\ref{formal-Pint}) in which only
the closest BH is taken into account.
In this case, the computation becomes much more feasible.
We found that the quantity $\alpha$ constructed from the merger rate density $\mathcal{R}$
in the BH mass plane as
\be
\alpha (m_1,m_2,t) \equiv -{(m_1+m_2)}^2 \frac{\partial^2}{\partial m_1 \partial m_2} \ln \mathcal{R}(m_1,m_2,t),
\ee
becomes almost independent of the PBH mass function and takes a value close to unity
($0.97 \lesssim \alpha \lesssim 1.05$).
Since it is possible that several distant BHs generate the dominant torque instead of the closest one during binary formation in the early universe, we have also considered the case
in which the remote BHs are taken into account for a flat PBH mass function.
Even in this case, we found that the quantity $\alpha$ exactly coincides with the one derived for the case of the closest perturbing BH.
This suggests that the determined value of $\alpha$ is robust to observationally test the PBH scenario once a large sample of mergers becomes available with accurately determined masses.

Other astrophysical mechanisms leading to BH mergers are generally expected to yield different $\alpha$ values. Recently, \citet{OLeary:2016ayz} has shown that the probability of merger is proportional to $m_{\rm t}^4$ for binary BH mergers in dense star clusters, which implies $\alpha\sim 4$ if the merger rates are nearly independent to mass ratio. 
PBH binaries formed in the low redshift Universe by GW emission during close encounters leads to $\alpha \approx 1.43$ \citep{Bird:2016dcv}. BH binaries formed by GW emission in mass-segregated environments such as galactic nuclei lead to $\alpha$ values that vary with the total binary mass $m_{\rm t}$ \citep{Gondanetal2017}.

The mass distribution is not the only 
GW observable which allows one to distinguish between different mechanisms leading to binary BH mergers.
For instance, it was shown recently that PBHs are unlikely to possess large spins \citep{Chiba:2017rvs}.
When the statistics of BH spins is accumulated in the future, 
this will also become a powerful discriminator. 
Further, the eccentricity distribution will be useful to distinguish binaries formed by GW capture in high velocity dispersion environments at low redshifts \citep{OLeary:2008myb,Gondan:2017hbp}. The observable PBH binaries that formed at high redshifts are expected to have close to zero eccentricity due to circularization by GW emission \citep{Peters:1964zz}. {\it LISA} will be able to determine the eccentricity for mergers with $e \gtrsim 10^{-6}$ \citep{Seto:2016wom}. 
Detection of BHs with masses less than $\sim 1\,\Msun$, which may be possible with
the advanced LIGO, VIRGO, and KAGRA at design sensitivity, would provide strong evidence of the existence of PBHs \citep{Magee:2017vkk,Clesseetal2017}.
Finally, future GW detectors will allow us to map out the cosmological luminosity distance (or redshift) distribution 
for BH mergers to high redshifts \citep{Nakamura:2016hna,Koushiappas:2017kqm}.
Examining the multidimensional GW event rate distribution will be essential to prove or disprove the PBH scenario.
 
\acknowledgments{
This work was supported by MEXT KAKENHI Nos. 17H06357 (T.T. and T.S.), 17H06358 (T.T.),
17H06359 (T.S.), 15H05888 (T.S and S.Y.), 15H02087 (T.T.), 
and 15K21733 (T.S. and S.Y.),
JSPS Grant-in-Aid for Young Scientists (B) No.15K17632
(T.S.) and No.15K17659 (S.Y.), 
the Grant-in-Aid for Scientific Research No. 26287044 (T.T.).
This project has received funding from the European Research Council (ERC) under the European Union's Horizon 2020 research and innovation programme under grant agreement No 638435 (GalNUC) and by the Hungarian National
 Research, Development, and Innovation Office grant NKFIH KH-125675 (B.K.). This work was performed in 
 part at the Aspen Center for Physics, which is supported by National Science Foundation grant 
 PHY-1607761.
}

\appendix
\section{Derivation of the probability distribution}
\label{app-formal-F}
Non-trivial part of Eq.~(\ref{formal-F}) is the probability distribution
for $x$ and $y_i~(i=1,\cdots,N)$, and we focus on this part only.

\begin{figure}[tbp]
  \begin{center}
   \includegraphics[width=150mm]{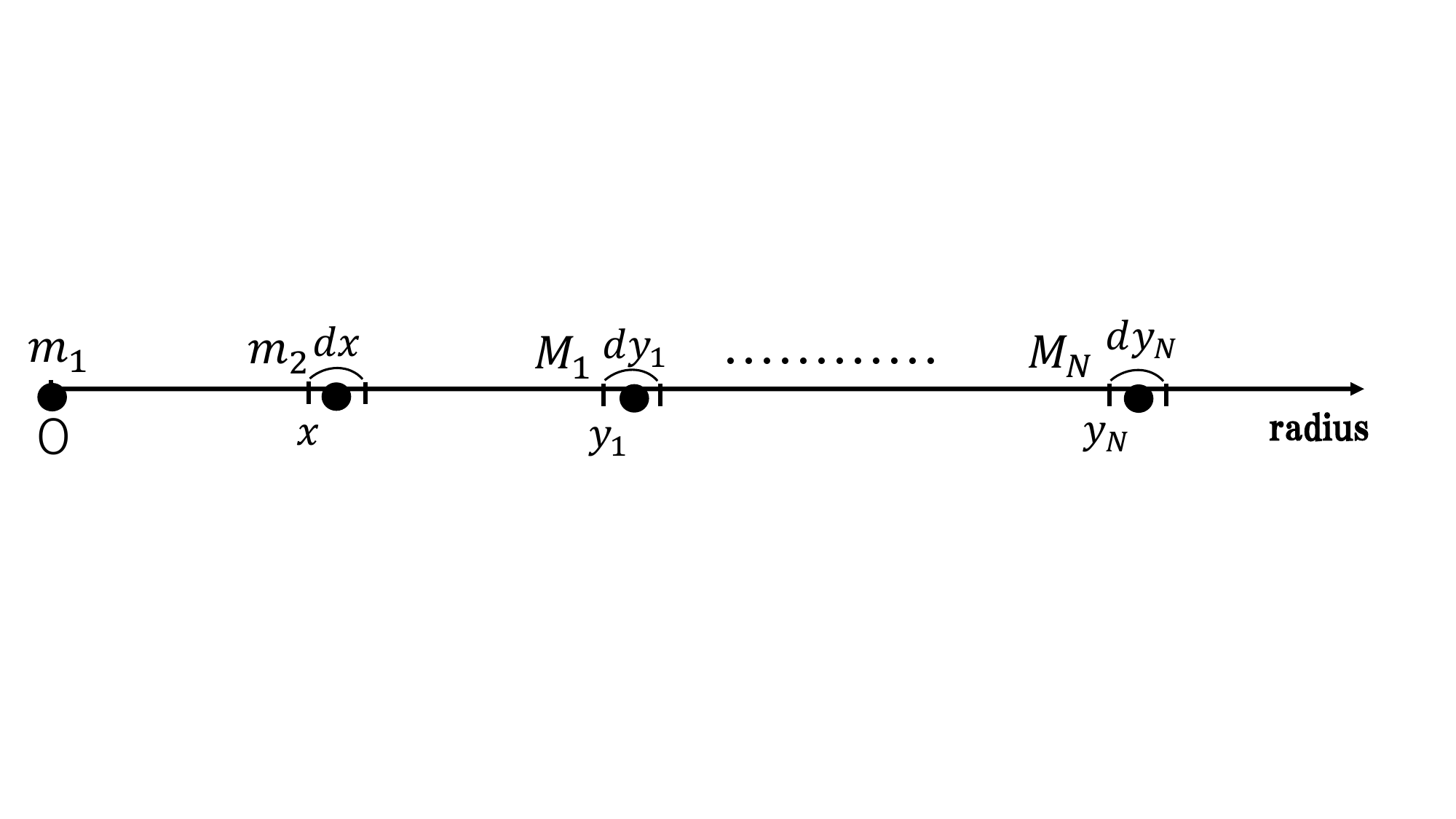}
  \end{center}
  \caption{This figure describes a situation where the individual BHs with mass
  $m_1,~m_2,~M_1,\cdots$ and $M_N$ locate at
  the origin, $(x,x+dx),~(y_1,y_1+dy_1), \cdots,$ and $(y_N,y_N+dy_N)$, respectively.}
  \label{appendix}
\end{figure}

Let $P(N,V)$ be the probability that there are $N$ BHs in the volume $V$.
For BHs that are uniform randomly distributed, we have
\be
P(N,V)=\frac{1}{N!} {\left( \frac{V}{V_0} \right)}^N e^{-V/V_0},
\ee
where $V_0$ is the volume for which the expectation particle number is 1.
Thus,
\be
V_0=n_{\rm BH}^{-1}.
\ee
Then, the probability that the situation shown in Fig.~\ref{appendix} is realized is given by
\begin{align}
dP&=P\left( 0,\frac{4\pi}{3} x^3 \right) \frac{d\left( \frac{4\pi}{3} x^3 \right)}{V_0}
P\left( 0, \frac{4\pi}{3} y_1^3-\frac{4\pi}{3} x^3 \right)  \frac{d\left( \frac{4\pi}{3} y_1^3 \right)}{V_0}
\cdots P\left( 0, \frac{4\pi}{3} y_N^3-\frac{4\pi}{3} y_{N-1}^3 \right) \frac{d\left( \frac{4\pi}{3} y_N^3 \right)}{V_0} \nonumber \\
&=\frac{4\pi x^2 dx}{V_0} \frac{4\pi y_1^2 dy_1}{V_0} \cdots \frac{4\pi y_N^2 dy_N}{V_0} 
\exp \left( -\frac{4\pi y_N^3}{3 V_0} \right).
\end{align}

\label{app-flat-variance}
From the definition of ${\vec \zeta}$ in Eq.~(\ref{eccen-e}),
we have
\be
\langle {\vec \zeta}^2 \rangle=\frac{x^6}{m_{\rm t}^2} \sum_{i=1}^N \sum_{j=1}^N
\bigg\langle \frac{1}{y_i^3}\frac{1}{y_j^3} \bigg\rangle \langle M_i M_j \rangle
\bigg\langle \sin (2\theta_i) \sin (2\theta_j)
\frac{({\vec e_z}\times {\vec e_i})}{| {\vec e_z}\times {\vec e_i} |} 
\cdot \frac{({\vec e_z}\times {\vec e_j})}{| {\vec e_z}\times {\vec e_j} |} \bigg\rangle.
\ee
Using Eq.~(\ref{unit-vector-ei}) for ${\vec e_i}$, we obtain
\be
\bigg\langle \sin (2\theta_i) \sin (2\theta_j)
\frac{({\vec e_z}\times {\vec e_i})}{| {\vec e_z}\times {\vec e_i} |} 
\cdot \frac{({\vec e_z}\times {\vec e_j})}{| {\vec e_z}\times {\vec e_j} |} \bigg\rangle 
= \frac{8}{15} \delta_{ij}.
\ee
By the assumption that $M_i$ obeys the uniform distribution in the interval $(\epsilon m_{\rm max},m_{\rm max})$,
we have
\be
\langle M_i^2 \rangle = \frac{1}{3} m_{\rm max}^2 (1+\epsilon+\epsilon^2).
\ee
Thus, we obtain
\be
\langle {\vec \zeta}^2 \rangle=\frac{8}{45} \frac{x^6}{m_{\rm t}^2} m_{\rm max}^2 (1+\epsilon+\epsilon^2)
\sum_{i=1} \bigg\langle \frac{1}{y_i^6} \bigg\rangle. \label{app2}
\ee
The calculation of $\sum_{i=1}\langle 1/y_i^6 \rangle$ can be done by noting that 
it is an expectation value of $1/y^6$ where $y$ is the distance of 
particles randomly distributed in the region $y >x$ \citep{Ioka:1998nz}, 
\be
\lim_{N\to \infty} \sum_{i=1}^N \bigg\langle \frac{1}{y_i^6} \bigg\rangle = \int_x^\infty \frac{4\pi y^2 dy}{n_{\rm BH}^{-1}} \frac{1}{y^6}=
\frac{4\pi}{3} \frac{n_{\rm BH}}{x^3}.
\ee
Plugging this result into Eq.~(\ref{app2}) finally yields
\be
\langle {\vec \zeta}^2 \rangle=\frac{32\pi}{135} n_{\rm BH} x^3 {\left( \frac{m_{\rm max}}{m_{\rm t}} \right)}^2 (1+\epsilon+\epsilon^2).
\ee

\bibliographystyle{yahapj}
\bibliography{draft}

\begin{thebibliography}{}
\providecommand\natexlab[1]{#1}
\providecommand\JournalTitle[1]{#1}

\bibitem[{Abbott {et~al.}(2016{\natexlab{a}})}]{TheLIGOScientific:2016htt}
Abbott, B.~P., {et~al.} 2016{\natexlab{a}},
  \href{http://dx.doi.org/10.3847/2041-8205/818/2/L22}{\JournalTitle{Astrophys.
  J.}, 818, L22}

\bibitem[{Abbott {et~al.}(2016{\natexlab{b}})}]{TheLIGOScientific:2016pea}
---. 2016{\natexlab{b}},
  \href{http://dx.doi.org/10.1103/PhysRevX.6.041015}{\JournalTitle{Phys. Rev.},
  X6, 041015}

\bibitem[{Abbott {et~al.}(2016{\natexlab{c}})}]{Abbott:2016blz}
---. 2016{\natexlab{c}},
  \href{http://dx.doi.org/10.1103/PhysRevLett.116.061102}{\JournalTitle{Phys.
  Rev. Lett.}, 116, 061102}

\bibitem[{Abbott {et~al.}(2016{\natexlab{d}})}]{Abbott:2016nhf}
---. 2016{\natexlab{d}},
  \href{http://dx.doi.org/10.3847/2041-8205/833/1/L1}{\JournalTitle{Astrophys.
  J.}, 833, L1}

\bibitem[{Abbott {et~al.}(2017{\natexlab{a}})}]{Abbott:2017vtc}
---. 2017{\natexlab{a}},
  \href{http://dx.doi.org/10.1103/PhysRevLett.118.221101}{\JournalTitle{Phys.
  Rev. Lett.}, 118, 221101}

\bibitem[{Abbott {et~al.}(2017{\natexlab{b}})}]{Abbott:2017gyy}
---. 2017{\natexlab{b}},
  \href{http://dx.doi.org/10.3847/2041-8213/aa9f0c}{\JournalTitle{Astrophys.
  J.}, 851, L35}

\bibitem[{Abbott {et~al.}(2017{\natexlab{c}})}]{Abbott:2017oio}
---. 2017{\natexlab{c}}, \JournalTitle{Submitted to: Phys. Rev. Lett.},
  \href{http://arxiv.org/abs/1709.09660}{{\sffamily arXiv:1709.09660 [gr-qc]}}

\bibitem[{Ali-Haimoud \& Kamionkowski(2017)}]{Ali-Haimoud:2016mbv}
Ali-Haimoud, Y., \& Kamionkowski, M. 2017,
  \href{http://dx.doi.org/10.1103/PhysRevD.95.043534}{\JournalTitle{Phys.
  Rev.}, D95, 043534}

\bibitem[{Ali-Haïmoud {et~al.}(2017)Ali-Haïmoud, Kovetz, \&
  Kamionkowski}]{Ali-Haimoud:2017rtz}
Ali-Haïmoud, Y., Kovetz, E.~D., \& Kamionkowski, M. 2017,
  \href{http://dx.doi.org/10.1103/PhysRevD.96.123523}{\JournalTitle{Phys.
  Rev.}, D96, 123523}

\bibitem[{Bird {et~al.}(2016)Bird, Cholis, Muñoz, Ali-Haimoud, Kamionkowski,
  Kovetz, Raccanelli, \& Riess}]{Bird:2016dcv}
Bird, S., Cholis, I., Muñoz, J.~B., {et~al.} 2016,
  \href{http://dx.doi.org/10.1103/PhysRevLett.116.201301}{\JournalTitle{Phys.
  Rev. Lett.}, 116, 201301}

\bibitem[{{Brandt}(2016)}]{Brandt:2016}
{Brandt}, T.~D. 2016,
  \href{http://dx.doi.org/10.3847/2041-8205/824/2/L31}{\JournalTitle{\apjl},
  824, L31}

\bibitem[{Byrnes {et~al.}(2012)Byrnes, Copeland, \& Green}]{Byrnes:2012yx}
Byrnes, C.~T., Copeland, E.~J., \& Green, A.~M. 2012,
  \href{http://dx.doi.org/10.1103/PhysRevD.86.043512}{\JournalTitle{Phys.
  Rev.}, D86, 043512}

\bibitem[{Carr {et~al.}(2016)Carr, Kuhnel, \& Sandstad}]{Carr:2016drx}
Carr, B., Kuhnel, F., \& Sandstad, M. 2016,
  \href{http://dx.doi.org/10.1103/PhysRevD.94.083504}{\JournalTitle{Phys.
  Rev.}, D94, 083504}

\bibitem[{Carr {et~al.}(2017)Carr, Raidal, Tenkanen, Vaskonen, \&
  Veermäe}]{Carr:2017jsz}
Carr, B., Raidal, M., Tenkanen, T., Vaskonen, V., \& Veermäe, H. 2017,
  \href{http://dx.doi.org/10.1103/PhysRevD.96.023514}{\JournalTitle{Phys.
  Rev.}, D96, 023514}

\bibitem[{Carr(1975)}]{Carr:1975qj}
Carr, B.~J. 1975,
  \href{http://dx.doi.org/10.1086/153853}{\JournalTitle{Astrophys. J.}, 201, 1}

\bibitem[{Carr(2005)}]{Carr:2005zd}
Carr, B.~J. 2005, in {59th Yamada Conference on Inflating Horizon of Particle
  Astrophysics and Cosmology Tokyo, Japan, June 20-24, 2005}

\bibitem[{Carr \& Hawking(1974)}]{Carr:1974nx}
Carr, B.~J., \& Hawking, S.~W. 1974, \JournalTitle{Mon. Not. Roy. Astron.
  Soc.}, 168, 399

\bibitem[{Chiba \& Yokoyama(2017)}]{Chiba:2017rvs}
Chiba, T., \& Yokoyama, S. 2017,
  \href{http://dx.doi.org/10.1093/ptep/ptx087}{\JournalTitle{PTEP}, 2017,
  083E01}

\bibitem[{Choptuik(1993)}]{Choptuik:1992jv}
Choptuik, M.~W. 1993,
  \href{http://dx.doi.org/10.1103/PhysRevLett.70.9}{\JournalTitle{Phys. Rev.
  Lett.}, 70, 9}

\bibitem[{{Clesse} \& {Garc{\'{\i}}a-Bellido}(2017)}]{Clesseetal2017}
{Clesse}, S., \& {Garc{\'{\i}}a-Bellido}, J. 2017, \JournalTitle{ArXiv
  e-prints}, \href{http://arxiv.org/abs/1711.10458}{{\sffamily
  arXiv:1711.10458}}

\bibitem[{Clesse \& Garc\'ia-Bellido(2017)}]{Clesse:2016vqa}
Clesse, S., \& Garc\'ia-Bellido, J. 2017,
  \href{http://dx.doi.org/10.1016/j.dark.2016.10.002}{\JournalTitle{Phys. Dark
  Univ.}, 15, 142}

\bibitem[{Custodio \& Horvath(1998)}]{Custodio:1998pv}
Custodio, P.~S., \& Horvath, J.~E. 1998,
  \href{http://dx.doi.org/10.1103/PhysRevD.58.023504}{\JournalTitle{Phys.
  Rev.}, D58, 023504}

\bibitem[{Eroshenko(2016)}]{Eroshenko:2016hmn}
Eroshenko, {\relax Yu}.~N. 2016,
  \href{http://arxiv.org/abs/1604.04932}{{\sffamily arXiv:1604.04932
  [astro-ph.CO]}}

\bibitem[{Fishbach \& Holz(2017)}]{Fishbach:2017zga}
Fishbach, M., \& Holz, D.~E. 2017,
  \href{http://dx.doi.org/10.3847/2041-8213/aa9bf6}{\JournalTitle{Astrophys.
  J.}, 851, L25}

\bibitem[{Gaggero {et~al.}(2017)Gaggero, Bertone, Calore, Connors, Lovell,
  Markoff, \& Storm}]{Gaggero:2016dpq}
Gaggero, D., Bertone, G., Calore, F., {et~al.} 2017,
  \href{http://dx.doi.org/10.1103/PhysRevLett.118.241101}{\JournalTitle{Phys.
  Rev. Lett.}, 118, 241101}

\bibitem[{{Ghosh} {et~al.}(2016){Ghosh}, {Del Pozzo}, \&
  {Ajith}}]{Ghoshetal2016}
{Ghosh}, A., {Del Pozzo}, W., \& {Ajith}, P. 2016,
  \href{http://dx.doi.org/10.1103/PhysRevD.94.104070}{\JournalTitle{\prd}, 94,
  104070}

\bibitem[{Gond\'an {et~al.}(2017)Gond\'an, Kocsis, Raffai, \&
  Frei}]{Gondan:2017hbp}
Gond\'an, L., Kocsis, B., Raffai, P., \& Frei, Z. 2017,
  \href{http://arxiv.org/abs/1705.10781}{{\sffamily arXiv:1705.10781
  [astro-ph.HE]}}

\bibitem[{{Gond{\'a}n} {et~al.}(2017){Gond{\'a}n}, {Kocsis}, {Raffai}, \&
  {Frei}}]{Gondanetal2017}
{Gond{\'a}n}, L., {Kocsis}, B., {Raffai}, P., \& {Frei}, Z. 2017,
  \JournalTitle{ArXiv e-prints},
  \href{http://arxiv.org/abs/1711.09989}{{\sffamily arXiv:1711.09989
  [astro-ph.HE]}}

\bibitem[{Green(2017)}]{Green:2017qoa}
Green, A.~M. 2017,
  \href{http://dx.doi.org/10.1103/PhysRevD.96.043020}{\JournalTitle{Phys.
  Rev.}, D96, 043020}

\bibitem[{Harada {et~al.}(2013)Harada, Yoo, \& Kohri}]{Harada:2013epa}
Harada, T., Yoo, C.-M., \& Kohri, K. 2013,
  \href{http://dx.doi.org/10.1103/PhysRevD.88.084051,
  10.1103/PhysRevD.89.029903}{\JournalTitle{Phys. Rev.}, D88, 084051},
  [Erratum: Phys. Rev.D89,no.2,029903(2014)]

\bibitem[{Hayasaki {et~al.}(2016)Hayasaki, Takahashi, Sendouda, \&
  Nagataki}]{Hayasaki:2009ug}
Hayasaki, K., Takahashi, K., Sendouda, Y., \& Nagataki, S. 2016,
  \href{http://dx.doi.org/10.1093/pasj/psw065}{\JournalTitle{Publ. Astron. Soc.
  Jap.}, 68, 66}

\bibitem[{Horowitz(2016)}]{Horowitz:2016lib}
Horowitz, B. 2016, \href{http://arxiv.org/abs/1612.07264}{{\sffamily
  arXiv:1612.07264 [astro-ph.CO]}}

\bibitem[{Inoue \& Kusenko(2017)}]{Inoue:2017csr}
Inoue, Y., \& Kusenko, A. 2017,
  \href{http://dx.doi.org/10.1088/1475-7516/2017/10/034}{\JournalTitle{JCAP},
  1710, 034}

\bibitem[{Ioka {et~al.}(1998)Ioka, Chiba, Tanaka, \& Nakamura}]{Ioka:1998nz}
Ioka, K., Chiba, T., Tanaka, T., \& Nakamura, T. 1998,
  \href{http://dx.doi.org/10.1103/PhysRevD.58.063003}{\JournalTitle{Phys.
  Rev.}, D58, 063003}

\bibitem[{Kashlinsky(2016)}]{Kashlinsky:2016sdv}
Kashlinsky, A. 2016,
  \href{http://dx.doi.org/10.3847/2041-8205/823/2/L25}{\JournalTitle{Astrophys.
  J.}, 823, L25}

\bibitem[{Koushiappas \& Loeb(2017{\natexlab{a}})}]{Koushiappas:2017chw}
Koushiappas, S.~M., \& Loeb, A. 2017{\natexlab{a}},
  \href{http://dx.doi.org/10.1103/PhysRevLett.119.041102}{\JournalTitle{Phys.
  Rev. Lett.}, 119, 041102}

\bibitem[{Koushiappas \& Loeb(2017{\natexlab{b}})}]{Koushiappas:2017kqm}
---. 2017{\natexlab{b}},
  \href{http://dx.doi.org/10.1103/PhysRevLett.119.221104}{\JournalTitle{Phys.
  Rev. Lett.}, 119, 221104}

\bibitem[{{Kovetz} {et~al.}(2017){Kovetz}, {Cholis}, {Breysse}, \&
  {Kamionkowski}}]{Kovetz:2017}
{Kovetz}, E.~D., {Cholis}, I., {Breysse}, P.~C., \& {Kamionkowski}, M. 2017,
  \href{http://dx.doi.org/10.1103/PhysRevD.95.103010}{\JournalTitle{\prd}, 95,
  103010}

\bibitem[{Magee \& Hanna(2017)}]{Magee:2017vkk}
Magee, R., \& Hanna, C. 2017,
  \href{http://dx.doi.org/10.3847/2041-8213/aa831c}{\JournalTitle{Astrophys.
  J.}, 845, L13}

\bibitem[{Mandel {et~al.}(2017)Mandel, Farr, Colonna, Stevenson, Tiňo, \&
  Veitch}]{Mandel:2016prl}
Mandel, I., Farr, W.~M., Colonna, A., {et~al.} 2017,
  \href{http://dx.doi.org/10.1093/mnras/stw2883}{\JournalTitle{Mon. Not. Roy.
  Astron. Soc.}, 465, 3254}

\bibitem[{Matsumoto {et~al.}(2017)Matsumoto, Teraki, \&
  Ioka}]{Matsumoto:2017adh}
Matsumoto, T., Teraki, Y., \& Ioka, K. 2017,
  \href{http://arxiv.org/abs/1704.05047}{{\sffamily arXiv:1704.05047
  [astro-ph.HE]}}

\bibitem[{Miller(2016)}]{Miller:2016krr}
Miller, M.~C. 2016,
  \href{http://dx.doi.org/10.1007/s10714-016-2088-4}{\JournalTitle{Gen. Rel.
  Grav.}, 48, 95}

\bibitem[{Nakamura {et~al.}(1997)Nakamura, Sasaki, Tanaka, \&
  Thorne}]{Nakamura:1997sm}
Nakamura, T., Sasaki, M., Tanaka, T., \& Thorne, K.~S. 1997,
  \href{http://dx.doi.org/10.1086/310886}{\JournalTitle{Astrophys. J.}, 487,
  L139}

\bibitem[{Nakamura {et~al.}(2016)}]{Nakamura:2016hna}
Nakamura, T., {et~al.} 2016,
  \href{http://dx.doi.org/10.1093/ptep/ptw127}{\JournalTitle{PTEP}, 2016,
  093E01}

\bibitem[{{Naoz}(2016)}]{Naoz:2016}
{Naoz}, S. 2016,
  \href{http://dx.doi.org/10.1146/annurev-astro-081915-023315}{\JournalTitle{\araa},
  54, 441}

\bibitem[{Niemeyer \& Jedamzik(1998)}]{Niemeyer:1997mt}
Niemeyer, J.~C., \& Jedamzik, K. 1998,
  \href{http://dx.doi.org/10.1103/PhysRevLett.80.5481}{\JournalTitle{Phys. Rev.
  Lett.}, 80, 5481}

\bibitem[{O'Leary {et~al.}(2009)O'Leary, Kocsis, \& Loeb}]{OLeary:2008myb}
O'Leary, R.~M., Kocsis, B., \& Loeb, A. 2009,
  \href{http://dx.doi.org/10.1111/j.1365-2966.2009.14653.x}{\JournalTitle{Mon.
  Not. Roy. Astron. Soc.}, 395, 2127}

\bibitem[{O'Leary {et~al.}(2016)O'Leary, Meiron, \& Kocsis}]{OLeary:2016ayz}
O'Leary, R.~M., Meiron, Y., \& Kocsis, B. 2016,
  \href{http://dx.doi.org/10.3847/2041-8205/824/1/L12}{\JournalTitle{Astrophys.
  J.}, 824, L12}

\bibitem[{Peters(1964)}]{Peters:1964zz}
Peters, P.~C. 1964,
  \href{http://dx.doi.org/10.1103/PhysRev.136.B1224}{\JournalTitle{Phys. Rev.},
  136, B1224}

\bibitem[{Poulin {et~al.}(2017)Poulin, Serpico, Calore, Clesse, \&
  Kohri}]{Poulin:2017bwe}
Poulin, V., Serpico, P.~D., Calore, F., Clesse, S., \& Kohri, K. 2017,
  \href{http://dx.doi.org/10.1103/PhysRevD.96.083524}{\JournalTitle{Phys.
  Rev.}, D96, 083524}

\bibitem[{Raidal {et~al.}(2017)Raidal, Vaskonen, \& Veermäe}]{Raidal:2017mfl}
Raidal, M., Vaskonen, V., \& Veermäe, H. 2017,
  \href{http://dx.doi.org/10.1088/1475-7516/2017/09/037}{\JournalTitle{JCAP},
  1709, 037}

\bibitem[{Sasaki {et~al.}(2016)Sasaki, Suyama, Tanaka, \&
  Yokoyama}]{Sasaki:2016jop}
Sasaki, M., Suyama, T., Tanaka, T., \& Yokoyama, S. 2016,
  \href{http://dx.doi.org/10.1103/PhysRevLett.117.061101}{\JournalTitle{Phys.
  Rev. Lett.}, 117, 061101}

\bibitem[{Seto(2016)}]{Seto:2016wom}
Seto, N. 2016,
  \href{http://dx.doi.org/10.1093/mnrasl/slw060}{\JournalTitle{Mon. Not. Roy.
  Astron. Soc.}, 460, L1}

\bibitem[{{Vitale} {et~al.}(2017){Vitale}, {Lynch}, {Raymond}, {Sturani},
  {Veitch}, \& {Graff}}]{Vitaleetal2017}
{Vitale}, S., {Lynch}, R., {Raymond}, V., {et~al.} 2017,
  \href{http://dx.doi.org/10.1103/PhysRevD.95.064053}{\JournalTitle{\prd}, 95,
  064053}

\bibitem[{Yokoyama(1998)}]{Yokoyama:1998xd}
Yokoyama, J. 1998,
  \href{http://dx.doi.org/10.1103/PhysRevD.58.107502}{\JournalTitle{Phys.
  Rev.}, D58, 107502}

\bibitem[{Young \& Byrnes(2013)}]{Young:2013oia}
Young, S., \& Byrnes, C.~T. 2013,
  \href{http://dx.doi.org/10.1088/1475-7516/2013/08/052}{\JournalTitle{JCAP},
  1308, 052}

\bibitem[{Zevin {et~al.}(2017)Zevin, Pankow, Rodriguez, Sampson, Chase,
  Kalogera, \& Rasio}]{Zevin:2017evb}
Zevin, M., Pankow, C., Rodriguez, C.~L., {et~al.} 2017,
  \href{http://dx.doi.org/10.3847/1538-4357/aa8408}{\JournalTitle{Astrophys.
  J.}, 846, 82}

\end{thebibliography}

\end{document}